\documentclass[acmsmall]{acmart}
\usepackage{tabularx}
\usepackage{multirow}
\usepackage{multicol}
\usepackage{colortbl}
\AtBeginDocument{%
  }

\setcopyright{acmlicensed}
\copyrightyear{2024}
\acmYear{2024}
\acmDOI{XXXXXXX.XXXXXXX}

\acmConference[CSCW '25]{ACM Conference on Computer-Supported Cooperative Work and Social Computing}{}{}

\usepackage{dblfloatfix}

\usepackage{csquotes}
\usepackage{subcaption}
\begin{document}

\title[From Fake Perfects to Conversational Imperfects]{From Fake Perfects to Conversational Imperfects: Exploring Image-Generative AI as a Boundary Object for Participatory Design of Public Spaces}

\author{Jose A. Guridi}
\email{jg2222@cornell.edu}
\orcid{0000-0003-0543-699X}
\affiliation{%
  \institution{College of Computing and Information Science, Cornell University}
  \city{Ithaca}
  \state{New York}
  \country{USA}
}

\author{Angel Hsing-Chi Hwang}
\orcid{0000-0002-0951-7845}
\email{angel.hwang@usc.edu}
\affiliation{%
  \institution{School for Communication and Journalism, University of Southern California}
  \city{Los Angeles}
  \state{California}
  \country{USA}
}

\author{Duarte Santo}
\orcid{0000-0002-2065-0065}
\affiliation{%
  \institution{Independent Scholar}
  \city{Madeira Island}
  \country{Portugal}
}

\author{Maria Goula}
\orcid{0009-0007-8296-4637}
\affiliation{%
  \institution{College of Agriculture and Life Sciences, Cornell University}
  \city{Ithaca}
  \state{New York}
  \country{USA}
}

\author{Cristobal Cheyre}
\orcid{0000-0002-1221-1978}
\email{cac555@cornell.edu}
\affiliation{%
  \institution{College of Computing and Information Science, Cornell University}
  \city{Ithaca}
  \state{New York}
  \country{USA}
}

\author{Lee Humphreys}
\orcid{0000-0002-5005-0394}
\affiliation{%
  \institution{College of Agriculture and Life Sciences, Cornell University}
  \city{Ithaca}
  \state{New York}
  \country{USA}
}

\author{Marco Rangel}
\orcid{0009-0009-8149-413X}
\affiliation{%
  \institution{Studio-MLA}
  \city{Los Angeles}
  \state{California}
  \country{USA}
}

\renewcommand{\shortauthors}{Guridi et al.}

\begin{abstract}
Designing public spaces requires balancing the interests of diverse stakeholders within a constrained physical and institutional space. Designers usually approach these problems through participatory methods but struggle to incorporate diverse perspectives into design outputs. The growing capabilities of image-generative artificial intelligence (IGAI) could support participatory design. Prior work in leveraging IGAI's capabilities in design has focused on augmenting the experience and performance of individual creators. We study how IGAI could facilitate participatory processes when designing public spaces, a complex collaborative task. We conducted workshops and IGAI-mediated interviews in a real-world participatory process to upgrade a park in Los Angeles. We found (1) a shift from focusing on accuracy to fostering richer conversations as the desirable outcome of adopting IGAI in participatory design, (2) that IGAI promoted more space-aware conversations, and (3) that IGAI-mediated conversations are subject to the abilities of the facilitators in managing the interaction between themselves, the AI, and stakeholders. We contribute by discussing practical implications for using IGAI in participatory design, including success metrics, relevant skills, and asymmetries between designers and stakeholders. We finish by proposing a series of open research questions.
\end{abstract}



\received{02 July 2024}

\maketitle

\section{Introduction}
Designing public spaces is a complex process that requires balancing the interests of diverse stakeholders and conquering constraints from physical, cultural, and institutional spaces.
Designers generally rely on participatory methods to identify, comprehend, and critically assess diverse needs and preferences from multi-stakeholders~\cite[e.g., ][]{hou_community-driven_2003, shaftoe_convivial_2012, raaphorst_visualization_2019, gaete_cruz_towards_2023}. 
However, designers often struggle to successfully elicit how to represent different cultures and identities in the built environment due to existing asymmetries between themselves and different stakeholders, such as power dynamics, communication barriers, cultural sensibility, and access to resources and information~\cite{hou_transcultural_2013, delgado_participatory_2023}. 

To facilitate conversations with stakeholders, designers often rely on tools and methods such as reference images, generative urban design models, and different technological platforms~\cite[e.g., ][]{boudjelida_electronic_2016,quan_urban-gan_2022, ardhianto_generative_2023, eanes_participatory_2018, jiang_generative_2023,raaphorst_visualization_2019}. 
Despite such efforts, research has found that perspectives from various communities are still not equally incorporated into design outputs, often leaving out the needs of underrepresented groups and their cultural heritage~\cite{jiang_generative_2023, hou_transcultural_2013}.
With the rapidly growing capabilities of image-generative artificial intelligence (IGAI), recent work advocates the potential of leveraging IGAI tools to support communication and collaboration between designers and their public stakeholders~\cite{guridi_image_2024, jiang_generative_2023}.
We examine whether IGAI tools deliver their promise and identify issues that remain as obstacles to effective co-design processes~\cite{guridi_image_2024, zhang_text--image_2023, struppek_exploiting_2024, schramowski_safe_2023}.

Applying IGAI to support public engagement in designing public spaces opens up a new avenue for human-AI co-design.
While leveraging generative AI's capabilities in design tasks has attained extensive research interests, the majority of such work focuses on how AI-powered tools augment the experience and performance of \textit{individual creators}~\cite[e.g., ][]{chiou_designing_2023, koch_may_2019, chung_artinter_2023, hwang_too_2022, lawton_when_2023,wang_reprompt_2023}.
Commonly, individual creators have established preferences, targeted audiences, and clear expectations for how they might respond to new technologies and tools.
By contrast, designing public spaces demands more extensive, multi-stakeholder involvement, while each participant lacks awareness of the other's needs and wants.
Moreover, practitioners often encounter constraints from physical spaces and institutional hurdles. These challenges are less apparent to those without rich domain knowledge, making it more challenging to avoid design obstacles proactively.

To explore how IGAI could facilitate public space design, we conducted a case study using IGAI as a boundary object in interviews with young legal immigrants to elicit information for designing a public park in Los Angeles.
Our study focuses on how designers and stakeholders interact and how IGAI shaped their discussion of experiences, needs, and desired qualities to design a more welcoming public space. 
We investigate the following question: \textbf{How can IGAI support participatory processes when designers and stakeholders co-design public spaces?}
This study is part of a larger project that analyzes \textbf{how technology can help spatialize translocality by identifying immigrants' desires, visions, and tensions in cultural and spatial negotiation.} 

We conducted a three-stage study to explore, examine, and reflect on the use of IGAI tools as boundary objects to mediate conversations between designers and stakeholders in public space design.
We collected data and extracted insights from multiple interviews and workshops. We organize our findings into three themes as follows:
(1) how multi-stakeholders jointly redefine successful generative output in IGAI-mediated participatory sessions, 
(2) how IGAI promoted space-aware conversations, and 
(3) how IGAI-mediated conversations are subject to the abilities of the facilitators in managing the interaction between themselves, the AI, and stakeholders.
All these findings shift the focal point from ''\textit{Fake Perfects}'' to ''\textit{Conversational Imperfects},'' suggesting that producing 'imperfect' results that prompt richer conversations should be prioritized over attaining accurate ('perfect') images in participatory processes.

We discuss three core contributions:
First, we provide practical implications for designing technologies to seek public engagement in public space design. 
We discuss the need to shift from object- to process-oriented participatory methods; this implies future participatory tool design should focus on interfaces and features that enable moderators of a participatory session to probe richer conversations with their participants.
Second, we discuss the importance of AI literacy, technical expertise, and soft skills to navigate conversations when adopting novel technologies to facilitate participatory processes.
Third, we propose a research agenda for technology-mediated engagement in the design of public spaces. 
We encourage future work to investigate these open research questions.

\section{Related Work}

\subsection{Designing Public Spaces}
\label{sec:rw_public_space_design}

The design of public spaces is often considered a wicked problem\footnote{Wicked problems are social problems which are inherently ill-defined and rely on political judgment for resolution~\cite{rittel_dilemmas_1973}.} encompassing multiple dimensions \cite{churchman_guest_1967,rittel_dilemmas_1973}.
Spatial designers must address large-scale, complex environments that involve physical, social, and regulatory dimensions and multifunctional requirements derived from diverse stakeholders' needs, preferences, and expectations~\cite{jacobs_death_2011,alexander_pattern_2017,whyte_social_2010,lefebvre_production_2013,healey_collaborative_2006,talen_problem_2000,albrechts_strategic_2004}. 
To do so, landscape designers must collaborate across disciplines, navigating a web of regulatory constraints and dynamic needs~\cite{jacobs_toward_2015, gehl_cities_2010} with care and contextual sensitivity~\cite{carmona_public_2021, lynch_image_1996}.

When designing for public problems, participation is a core element in generating value for constituents~\cite{boyne_public_2002, rose_managing_2015, bonina_public_2009, persson_government_2010}. 
Participation can bring together diverse people, providing benefits such as access to experience-based knowledge, enhanced support and legitimacy, and improved inclusiveness, transparency, and accountability~\cite{aitamurto_value_2017, arana-catania_citizen_2021,aitamurto_unmasking_2017,aitamurto_disagreement_2023}.
However, participatory methods can also harm people when ill-implemented~\cite{frahm_fixing_2022,delgado_participatory_2023,irvin_citizen_2004}. 
Too often, participation only simulates care and engagement, having limited impact on the outcomes of the public process~\cite{Williams2022,delgado_participatory_2023} and, thus, staying as an unfulfilled promise~\cite{Lovbrand2011}.

Landscape designers often rely on participatory processes to encompass stakeholders' conflicting preferences. 
For example, space designers use methods such as design charrettes~ \cite{lennertz_charrette_2014}, design workshops~\cite{sanders_co-creation_2008}, site walks~\cite{kanstrup_design_2014,clarke_more-than-human_2018}, design games~\cite{brandt_designing_2006}, participatory mapping~\cite{corbett_using_2006}, public interventions and exhibitions~\cite{lydon_tactical_2015,sanoff_community_1999}, and visual preference surveys~\cite{nasar_environmental_1992} to involve communities in the design process, reflect on local and environmental considerations, and promote a sense of ownership~\cite{hester_design_2006}. 

Despite employing various methods, traditional participatory approaches may not always be sufficient to capture all stakeholders' nuanced and diverse perspectives. Designers often encounter challenges in effectively eliciting how different cultures and identities wish to be represented in the built environment~\cite{hou_transcultural_2013}. 
Neither treating all issues as design problems nor using participatory approaches are universal fixes~\cite{disalvo_design_2022,delgado_participatory_2023,frahm_fixing_2022}.
Participatory design should be implemented through a collective task of caring and taking action for our future~\cite{disalvo_design_2022,carmona_public_2021}, considering institutional constraints as a resource for new strategies~\cite{lodato_institutional_2018}. 
However, to do so, designers must deal with the challenge of the existing asymmetries between them and the public~\cite{hou_transcultural_2013, delgado_participatory_2023}, how publics are constructed (i.e., who and how participate) \cite{disalvo_design_2009,dantec_infrastructuring_2013}, and how objects can have a material agency and influence on the process~\cite{jenkins_object-oriented_2016}.

Many methods and tools described above can be conceptualized as boundary objects, which could bridge asymmetries between designers and stakeholders~\cite{leigh_star_chapter_1989, leigh_star_this_2010}. 
Star~\cite{leigh_star_chapter_1989} proposes three characteristics of boundary objects: they provide a common language between participants to share knowledge, enable individuals to learn their differences, and facilitate collaboration in transforming knowledge. 
Boundary objects in design can create friction and controversy to promote discussion and learn from differences~\cite{huybrechts_living_2009}, establish and destabilize protocols to push boundaries~\cite{lee_boundary_2007} and ground ideas and bridge language across different backgrounds and disciplines~\cite{balint_design_2017}.

\subsection{Generative AI as Design Support}

Generative AI creates new content based on user input; specifically, IGAI applications generate \textit{image content} as users insert text and/or prompts in other modalities (see Zhang et al.~\cite{zhang_text--image_2023} for a review). 
A growing body of work has recently explored IGAI's potential in supporting various design tasks, demonstrating how IGAI applications could be incorporated into individual creators' workflows.
The use of IGAI tools for design support can range from probing design questions and collecting relevant information~\cite{hwang_too_2022, shi_understanding_2023, chung_artinter_2023}, exploring different ideas~\cite{hwang_too_2022, yang_sketching_2019, rajcic_towards_2024, lu_designing_2021, wan_gancollage_2023, chiou_designing_2023}, to verifying the feasibility of design~\cite{shi_understanding_2023, hwang_too_2022, huang_plantography_2024, rajcic_towards_2024}.
However, designing public spaces grapples with a unique set of challenges (as mentioned in Section~\ref{sec:rw_public_space_design}), and the utility of IGAI tools in addressing them remains underexplored.

Some generative tools have been tested in areas of landscape architecture. 
~For example, in designing landscapes, Huang et al. \cite{huang_plantography_2024} developed PlantoGraphy, a design system that allows an interactive configuration of IGAI to transform conceptual scenes into realistic landscape renderings. 
However, the system has limited flexibility in accommodating more diverse design requirements, which is a relevant challenge when designing complex public spaces. 
Guridi et al.~\cite{guridi_image_2024} discussed how IGAI could increase efficiency, multiplicity, and stakeholder engagement in co-designing public parks. 
Jiang et al.~\cite{jiang_generative_2023} conducted a systematic review of the generative urban design field and found that models are yet to address the complexity of the required elements, focusing primarily on single-element design such as streets and buildings~\cite{lima_urban_2021, yuan_physical_2016}. 
However, most research focuses on the quantitative performance of the urban design (e.g., traffic, energy efficiency)~\cite{vasanthakumar_performance-based_2017, khallaf_performance-based_2017, jiang_generative_2023} or high-quality images \cite{huang_plantography_2024}, which can fail to promote divergent thinking and holistic design options.

A particular role we highlight is how IGAI can become a boundary object between designers and stakeholders. 
Boundary objects support translation between social worlds~\cite{leigh_star_this_2010, leigh_star_chapter_1989, lee_boundary_2007} seeking to arrive at a mutual understanding. 
However, boundary objects have their innate subjectivity and ambiguity, sometimes making their role in translation difficult, failing to satisfy the informational needs of the parties, or requiring additional explanations to be intelligible~\cite{lee_boundary_2007,subrahmanian_boundary_2003,henderson_line_1998}. 
Recent studies analyzed IGAI as boundary objects in art creation~\cite{chung_artinter_2023}, but no work is related to participatory processes or landscape design.

However, using IGAI to support the design process entails risks. 
IGAI could harm people through different mechanisms such as generating offensive content~\cite{schramowski_safe_2023}, producing content biased in cultural, racial, or other relevant dimensions~\cite{struppek_exploiting_2024, bansal_how_2022,qadri_ais_2023,vazquez_taxonomy_2024,zhou_bias_2024}, generating hard to identify fake content when the truth is expected~\cite{ricker_towards_2024,sha_-fake_2023, corvi_detection_2023}, plagiarism~\cite{ghosh_can_2022}, among others.

In terms of interaction, the HCI field has long studied how human-AI (HAII) interaction is challenging, especially involving language technologies~\cite[e.g., ][]{jung_using_2015, sebo_robots_2020, suh_ai_2021, yang_re-examining_2020, yang_sketching_2019}. 
The HAII depends on personal traits~\cite{esterwood_personality_2020}, highly varying between individuals~\cite{sebo_robots_2020}. 
However, in the field of tools to support design, little research has delved into the details of HAII, leaving users largely hands-off during the experience \cite{hwang_too_2022}. 
For example, Chiou et al.~\cite{chiou_designing_2023} analyze the communication challenge in the designer-IGAI interaction and propose using design frameworks to aid in prompting generation. 
The latter is important since research suggests that the effectiveness of co-design with AI rests on participants’ chosen approach to prompt creation and their responses to the AI’s suggestions~\cite{wadinambiarachchi_effects_2024}.
Thus, designers might experience changes in how they work when incorporating IGAI. 
Chiou et al.,~\cite{chiou_designing_2023} found that designers might experience a transition from a creator to a curator role. 
The latter could conflict with the actual adoption of the tools since designers could resist them if they lack agency or are not provided with explanations, instructions, or proper co-design frameworks~\cite{oh_i_2018,zhou_understanding_2024}.

\section{Methods}
This study is part of a broader project to understand how to map immigrant preferences and uses of public space through technology to foster inclusive design and management of the urban commons. We detailed all the stages and research activities conducted for this project in Table~\ref{tab:timeline}.
To carry out the entire project, we partnered with Studio MLA, a landscape architecture firm in Los Angeles, to test different methods in their participatory process for eliciting information from first and second-generation immigrants to inform the design of public parks.
We examined how IGAI could bridge designers and stakeholders when designing public spaces. 
In the following subsections, we describe the stages of our project, which was granted exempt status by our IRB.

\begin{table*}[h!]
    \centering
    \caption{Full timeline of the research project}
    \begin{tabular}{p{0.3\textwidth}|p{0.6\textwidth}}
    \toprule
       {\cellcolor{lightgray!50!}\textbf{Research Stage}} & {\cellcolor{lightgray!50!}\textbf{Research Activities}} \\
       \hline
       Broader Project & 1) Interviews with immigrants in Los Angeles in 2023. \\
       & 2) Mapping Workshop with immigrants in Los Angeles in 2023. \\
       & 3) Online interviews with immigrants in 2024. \\
       & 4) Workshop in Coney Island, New York in 2024. \\
       \hline
       \multicolumn{2}{l}{\cellcolor{lightgray!50!}Research stages and activities of the current study} \\
       \hline
       \multirow{3}{*}{Stage 1: Speculative Process} & 1) We used transcripts from the interviews done in 2023 to experiment with how IGAI could be used to sketch the design of public spaces.\\
       & 2) We conducted reflection workshops about using IGAI to sketch from interviews.\\
       & 3) We conducted internal interviews about the potential of IGAI for designing Public Spaces.\\
       \hline
       Stage 2: Piloting Process & We conducted pilot interviews using IGAI with researchers on our team.\\
       \hline
       Stage 3: IGAI-mediated interviews & 1) We conducted interviews using IGAI with immigrants in Los Angeles.\\
       & 2) We conducted a reflection workshop about the experience of using IGAI as a boundary object in interviews. \\
       \bottomrule
    \end{tabular}
    \label{tab:timeline}
\end{table*}

\subsection{Broader Project}
This study is part of a broader project examining how to improve participatory design methods for public spaces involving immigrants. 
To do so, we have conducted interviews and participatory workshops in Los Angeles and New York to test different methods for eliciting immigrants' memories, values, and desires for different landscape projects. 
In this paper, we focus on a portion of the universe of interviews and workshops, which are only conducted in Los Angeles related to improving two public parks.

The first set of interviews was conducted in 2023 in Los Angeles to elicit information about memories and cultural heritage.
These interviews aimed to analyze the information and see how memories and cultural heritage could inform the transformation of the Puente Hills Landfill into a park\footnote{Puente Hill Landfill is located near Los Angeles, California, and operated until October 31, 2013. The zone where the landfill is located is predominantly populated by immigrants (e.g., 70\% Latinx/Hispanic and 19\% Asian).}. 
Table \ref{tab:stage0} shows the participants' demographics of this stage.
Interviews in this stage elicited information about the public spaces that participants frequented and their use of social media.
We asked what type of public spaces they use and why, what they do in those public spaces, and how they use social media in relation to those spaces.
Moreover, we asked participants to think about the public parks they visited and the elements they valued and deemed missing.
Finally, we asked them about Puente Hill Landfill and the elements they wanted to see when it became a park.
During this process, we reflected on the challenge of overcoming asymmetries between designers and participants and decided to explore how IGAI could support the process. 

\begin{table}[h!]
    \centering
        \caption{Interviews from prior work with Immigrants in LA in 2023}
    \begin{tabular}{p{0.2\linewidth} p{0.2\linewidth} p{0.3\linewidth} p{0.2\linewidth}}
    \toprule
       \textbf{Participant} & \textbf{Gender} & \textbf{Country of Origin} & \textbf{Years in LA}\\
       \hline
       A1 & Male & Mexico & 21-25\\
       A2 & Male & El Salvador & 40+ \\
       A3 & Male & Mexico & 26-30  \\
       A4 & Male & El Salvador & 10-15 \\
       A5 & Male & Mexico & 0-5 \\
       A6 & Female & Mexico & 21-25 \\
       A7 & Female & Chile & 0-5\\
       A8 & Female & Chile & 0-5\\
    \bottomrule
    \end{tabular}
        \label{tab:stage0}
\end{table}

\subsection{Stage 1: Speculative Process}\label{stg1: speculative}
The first stage's goal was to explore the opportunities and challenges of using IGAI to support the design of public spaces. 
We assembled a team of students and professors in landscape architecture to use IGAI as a design probe~\cite{mattelmaki_design_2006} and reflect on opportunities and desires emerging from using IGAI in participatory design. 
During 6 months, the team worked on generating prompts from the first set of interviews (table~\ref{tab:stage0}). 
To do so, the team coded the transcripts using aspects of the Convivial Public Space Framework~\cite{shaftoe_convivial_2012, shedid_approach_2021}: Place, Activity, Features, Emotion, and Sense of Place.
Then, the team extracted keywords to develop the prompts and generated images of the parks described during the interviews, with and without reference images.
We include a sample of prompts and images from this prior work in annex~\ref{annex1}.
During the whole process, the team documented their experience.

To debrief the experience, we conducted four interviews with four team members (i.e., the researchers with a Landscape Architecture background) and organized a two-hour reflection workshop to discuss the main insights from using IGAI and the envisioned challenges of this method in real-time during the interviews. 
The main insights of this work served as the base for designing stages 2 and 3 and were published earlier in 2024~\cite{guridi_image_2024}. 
Table~\ref{tab:stage1} details the interviewees from the team. 

\begin{table}[h!]
    \centering
    \caption{Stage 1 Participants}
    \begin{tabular}{p{0.15\linewidth} p{0.6\linewidth} p{0.15\linewidth}}
    \toprule
       \multicolumn{3}{c}{\cellcolor{lightgray!20!}\textbf{Internal Interviews}}\\
       \hline
       \textbf{Participant} & \textbf{Education} & \textbf{Gender} \\
       \hline
       B1 & Ph.D. in Landscape Design Theory and Master in Landscape Architecture & Female\\
       B2 & Masters in Architecture \& Landscape Architecture & Male \\
       B3 & Master in Landscape Architecture & Female \\
       B4 & Architecture student & Male \\
    \bottomrule
    \end{tabular}
    
   \begin{tabular}{p{0.15\linewidth} p{0.6\linewidth} p{0.15\linewidth}}
    \toprule
       \multicolumn{3}{c}{\cellcolor{lightgray!20!}\textbf{Reflection Workshop}}\\
       \hline
       \textbf{Participant} & \textbf{Education} & \textbf{Gender} \\
       \hline
       C1 & Ph.D. in Landscape Design Theory and Master in Landscape Architecture & Female\\
       C2 & Masters in Architecture \& Landscape Architecture & Male \\
       C3 & Master in Landscape Architecture & Male \\
       C4 & Master in Landscape Architecture Student & Female \\
       C5 & Master in Landscape Architecture Student & Male \\
       C6 & Architecture student & Male \\
       C7 & Ph.D. in Information Science student & Male \\
       C8 & Ph.D. in Engineering and Public Policy & Male \\
       C9 & Ph.D. in Communication & Female \\
    \bottomrule
    \end{tabular}
        \label{tab:stage1}
\end{table}

\subsection{Stage 2: Piloting Process}
The second stage's goal was to design and test a protocol for using IGAI in a real-world intervention based on the insights from Stage 1. 
We conducted three internal interviews using IGAI to test the protocol, learn how to prompt and iterate in real-time, and reflect on the results, potential limitations, and challenges. 
To do so, the three research assistants who would prompt the system in the real-world experience interviewed three of the team's researchers.
During this stage, we focused only on the first section of the interview, about memories in a public space (see table~\ref{tab:interviews}). 
We analyzed the interviews to design the final protocol and to further reflect on the results of Stage 1.

\begin{table}[h!]
    \centering
        \caption{Stage 2 Participants}
   \begin{tabular}{p{0.15\linewidth} p{0.6\linewidth} p{0.15\linewidth}}
    \toprule
       \textbf{Participant} & \textbf{Education} & \textbf{Gender} \\
       \hline
       D1 & Ph.D. in Landscape Design Theory and Master in Landscape Architecture & Female\\
       D2 & Ph.D. in Information Science student & Male \\
       D3 & Ph.D. in Communication & Female \\
    \bottomrule
    \end{tabular}
    \label{tab:stage2}
\end{table}

\subsection{Stage 3: IGAI-mediated interviews}
The third stage aimed to test IGAI as boundary objects between designers and stakeholders.
We conducted nine IRB-approved semi-structured interviews using IGAI to generate images of participants' memories, desires, and requirements to upgrade River Garden Park in Los Angeles. 
For each participant, one researcher asked questions (facilitator), while a research assistant prompted the system to generate images based on the participant's responses (prompter).
We mainly used Dream Studio, but some specific tests were done with DALL-E 3, and Photoshop with Generative AI. Appendix~\ref{ap:software} provides a succinct description of each software.

The interview script was an adapted version of the one used during the 2023 interviews and described above in section~\ref{stg1: speculative}. 
We kept it as similar as possible, incorporating only the prototyping exercises with IGAI, modifying the section about Puente Hills Landfill to the River Garden Park, and deleting questions about social media that were no longer relevant. 
The interviews included three phases, and the researchers only moved forward when either the participant was satisfied with an image, no improvements were achieved after three iterations, or the interviewee had nothing more to say about the images. 
Table~\ref{tab:interviews} summarizes the details of the interview.
The goal of phase 1 was to illustrate a significant memory of the participant in a public park.
During this phase, the interviewees described an existing place and, alongside the facilitator and the prompter, tried to generate an illustration of that space and discuss their feelings about it.
The goal of Phase 2 was to discuss the public spaces that the participants often go to and, from that, co-design an ideal one for them.
The final phase consisted of presenting a set of pictures of River Garden Park and, from one that was picked by the interviewee, working together on upgrading the space.
Each interview was done in English and lasted approximately 60 minutes.

Participants were recruited by a nonprofit NGO working together with our industry partner to upgrade Rivergarden Park in Los Angeles.
All participants were second-generation immigrants from Latin America between 18 and 35 years old living in California and were compensated with USD \$40 in cash.
We recorded the audio and the computer screen in which we used the IGAI software.
Table~\ref{tab:stage3} summarizes participants' information.

\begin{table}[h!]
    \centering
        \caption{Stage 3 Participants}
    \begin{tabular}{p{0.3\linewidth} p{0.3\linewidth} p{0.3\linewidth}}
    \toprule
       \multicolumn{3}{c}{\cellcolor{lightgray!20!}\textbf{IGAI-mediated Interviews}}\\
       \textbf{Participant} & \textbf{Gender} & \textbf{Immigration origin}\\
       \hline
       E1 \& E2 \& E3 \& E4 \& E6 \& E7 & Female & Mexico\\
       E5 & Male & Mexico\\
       E8 & Female & Unknown\\
       E9 & Male & Guatemala\\
    \bottomrule
    \end{tabular}

   \begin{tabular}{p{0.15\linewidth} p{0.6\linewidth} p{0.15\linewidth}}
    \toprule
       \multicolumn{3}{c}{\cellcolor{lightgray!20!}\textbf{Reflection Workshop}}\\
       \hline
       \textbf{Participant} & \textbf{Education} & \textbf{Gender} \\
       \hline
       F1 & Ph.D. in Landscape Design Theory and Master in Landscape Architecture & Female\\
       F2 & Masters in Architecture \& Landscape Architecture & Male \\
       F3 & Master in Landscape Architecture & Male \\
       F4 & Master in Landscape Architecture Student & Female \\
       F5 & Undergraduate student with Major in Landscape Architecture & Male \\
       F6 & Master in Landscape Architecture Student & Male \\
       F7 & Architecture student & Male \\
       F8 & Ph.D. in Information Science student & Male \\
       F9 & Ph.D. in Engineering and Public Policy & Male \\
       F10 & Ph.D. in Communication & Female \\
    \bottomrule
    \end{tabular}
        \label{tab:stage3}
\end{table}

   

\begin{table*}[h!]
    \centering
    \caption{Interview Structure}
    \begin{tabular}{p{0.2\textwidth}|p{0.7\textwidth}}
    \toprule
       {\cellcolor{lightgray!50!}\textbf{Interview Phase}} & {\cellcolor{lightgray!50!}\textbf{Description}} \\
       \hline
       1) Memory & We asked participants to narrate a cherished memory in a public park. Then, we asked follow-up questions to elicit details and prompted the system to illustrate the memory.\\
       \midrule
       2) Generic public space & We asked participants about public spaces where they liked spending time. Then, we asked follow-up questions to elicit the features, environment, and sentiment participants had in mind and prompted the system to illustrate the space. One relevant focus of the questions was connecting the public space with the participants' cultural heritage.\\
       \midrule
       3) River Garden Park & Participants were shown a set of photographs of the River Garden Park and asked to choose one to work on. Then, we asked them how they would improve the park based on the picture, and we prompted the systems with that information. We used the responses from the first two stages to dig deeper into the details participants wished to include\\
       \bottomrule
    \end{tabular}
    \label{tab:interviews}
\end{table*}

After the interviews, we reflected on the experience through three means. 
First, each facilitator and prompter wrote an individual reflection on the experience. 
Second, the transcripts were analyzed alongside the images produced in the process. 
Third, based on preliminary insights, we organized a 2-hours workshop between the researchers and a representative of our industry partner, followed by a 1-hour wrap-up workshop a week later. 
In the workshops, we used a Miro board to discuss three elements: (1) The role of the images, (2) the interviewing process using IGAI, and the roles of the interviewer and interviewee. 
Finally, we used a combination of affinity diagrams~\cite{moggridge_designing_2007, shedid_approach_2021} and axial coding to analyze all the qualitative data.

\subsection{Positionality Statement}
Taking a ''researcher-as-instrument'' approach~\cite{researcher-as-instrument}, we built various research activities of the project upon the various expertise within our research team. 
We hereby provide a positionality statement to elaborate on the background and experiences of research team members.
We formed our research team with two landscape architecture researchers and practitioners, one communication researcher, one technology policy researcher, one digital technologies researcher, one HCI researcher, three landscape architecture research assistants, and one architecture research assistant.
The team's researchers have previously worked on diverse topics, such as landscape design, technology policy, human-AI interaction, and social media.
Moreover, our researchers have practical experience designing public spaces, implementing participatory processes, and working in the public and private sectors.
Table~\ref{tab:positionality} summarizes our team's self-identified demographics.
We also acknowledge that our findings are subject to biases related to our backgrounds and experiences when interpreting our participants' responses.
Nevertheless, we expect our research team members' various backgrounds to contribute to more diverse perspectives throughout the research process and acknowledge biases through deliberation and contrasting heterogeneous backgrounds.

\begin{table}[h!]
    \centering
        \caption{Research Team Demographics}
   \begin{tabular}{p{0.3\linewidth} p{0.3\linewidth} p{0.3\linewidth}}
    \toprule
       \textbf{Role} & \textbf{Country of Origin} & \textbf{Gender} \\
       \hline
       Author & Chile & Male\\
       Author & Taiwan & Female \\
       Author & Portugal & Male \\
       Author & Greece & Female \\
       Author & Chile & Male \\
       Author & USA & Female \\
       Author & USA & Non-binary \\
       Research Assistant & Mexico & Male \\
       Research Assistant & Ecuador & Male \\
       Research Assistant & China & Female \\ 
       Research Assistant & Greece & Male \\
       Research Assistant & China & Female \\ 
       Research Assistant & India & Female \\
    \bottomrule
    \end{tabular}
    \label{tab:positionality}
\end{table}

\section{Findings}
We conceptualize two types of IGAI-generated images in the co-design process. 
\textbf{"Fake Perfects"} were images that, although having high quality and/or text-image alignment, did not promote rich conversations. 
On the other hand, \textbf{"Conversational Imperfects"} were images that promoted rich conversations between practitioners and stakeholders despite having low text-image alignment or presenting imperfections.
The proposed categories are not exhaustive, but we propose them as analytical concepts from our findings.
From our case, we extract three main findings analyzing the dynamics of the Fake Perfects and Conversational Imperfects: 
(1) Success in IGAI is resignified from accurate to imperfect, conversational images, 
(2) 'Conversational Imperfects' promoted space-aware conversations, and 
(3) generative results are subject to practitioners managing reactions and interactions with the IGAI.

The following subsections explain our findings and provide evidence using images from the IGAI-mediated interviews and quotes from the interviews and workshops of the stages described in table~\ref{tab:timeline}. 
Participants are identified in accordance to the information in tables~\ref{tab:stage0},~\ref{tab:stage1},~\ref{tab:stage2} and~\ref{tab:stage3}. 
The phases of the IGAI-mediated interviews are summarized in table~\ref{tab:interviews}.
We use the following terminology to refer to the different groups. "Participants" are people recruited and interviewed; "facilitators" are researchers guiding the interviews; "prompters" are research assistants using the IGAI during interviews; and "practitioners" are researchers, research assistants, and people from our partnering firm with landscape architecture background.

\subsection{Finding 1: Success in IGAI is resignified from accurate to imperfect, conversational images}
Consistent with the existing literature, practitioners acknowledged the iterative nature of working with IGAI during the design processes.
However, practitioners who experienced frustration during stages 1 and 2 changed their minds when using IGAI in a real-world context in stage 3.
\textbf{Practitioners identified the benefits of those imperfect, intermediate outputs during the iterative processes of collaborating \textit{through} image generation.}

These distinct experiences could be attributed to their goals when IGAI was used as a medium for collaboration.
In alignment with the current prompting convention~\cite{zhang_text--image_2023}, the goals of practitioners in stages 1 and 2 focused on creating ''accurate images'' to represent and communicate their visions precisely.
This focus on accuracy often yielded frustration, and unexpected outputs were viewed as malperformance of IGAI or lack of capacities from the prompter.  
As researchers suggested: \blockquote[D1]{\textit{It took at least four or five changes to get a satisfactory picture. Sometimes, we felt a later picture was even worse than the first one.}} and \blockquote[D2]{\textit{There is a limited range of manipulation of images with prompts, potentially leading to significant differences [from expectations] and omissions of key elements}}.

In stark contrast, when generating images to mediate the conversation in stage 3, practitioners held distant opinions toward what were deemed as successful outcomes of prompting and working iteratively with IGAI.
In particular, we distilled two dynamics in which Conversational Imperfects (i.e., \textit{'imperfect'} images) were successful: (1) revealing hidden or unknown ideas and (2) inspiring new ideas. 

\subsubsection{Revealing hidden or unknown ideas}
\begin{figure*}[h!]
\begin{subfigure}[c]{0.45\textwidth}
  \caption{Park with a mural which recalled childhood memories from participant E2}
  \label{fig:mural}
    \centering
    \includegraphics[width = 0.9\columnwidth]{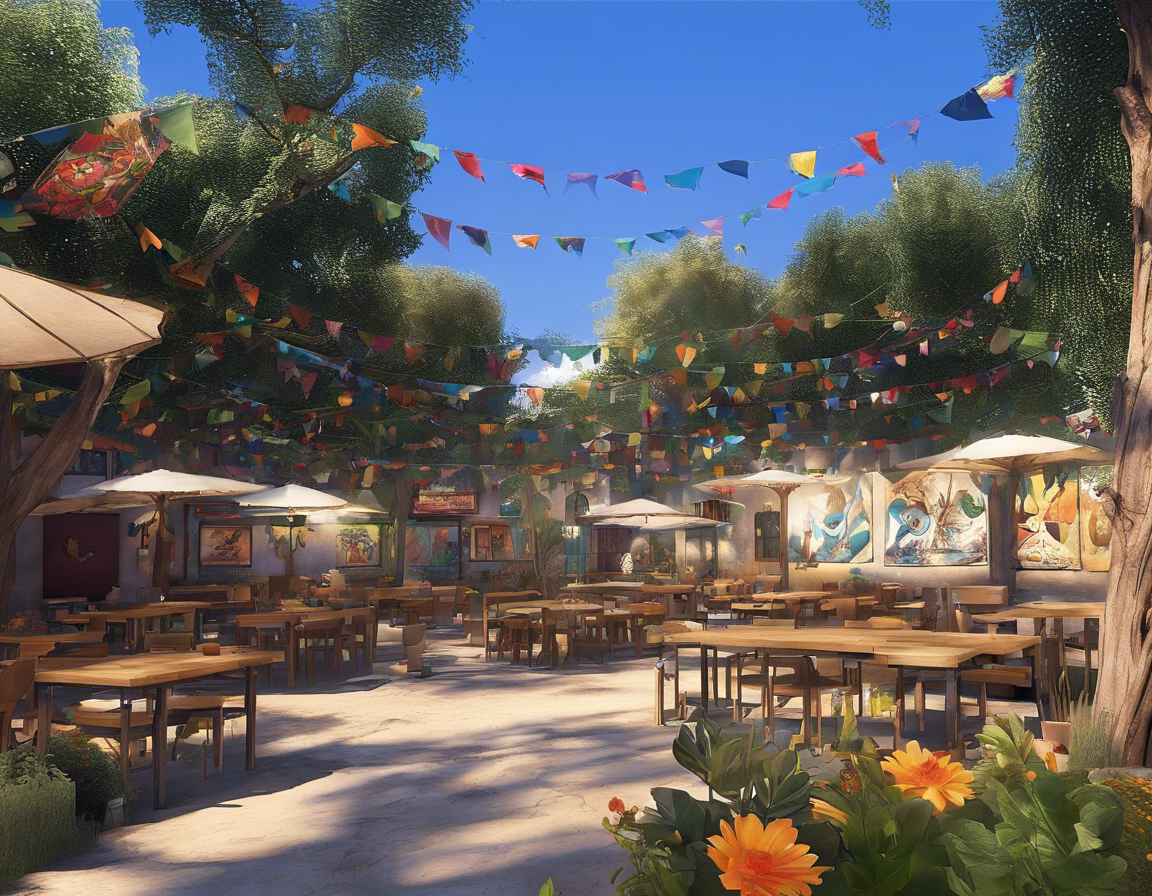}
      \caption*{\textbf{Prompt on Dream Studio}: Add a recreational area, add people relaxing, add food vendors, also include Mexican cultural paintings, a seating area, and flowers. \\
      \textbf{Reference image strength\footnote{Controls the amount of noise added to the input image. Higher values allow more variations, but the output might not be semantically consistent with the input~\cite{esser_stable_2022}}:} 27\%}
    \Description{The image depicts a square with wooden tables and small triangular flags of different colors hanging from cables. In the background, there are many trees and a mural.}
\end{subfigure}
\hfill
\begin{subfigure}[c]{0.45\textwidth}
  \caption{Alley with cacti alongside pink and orange flowers as discussed with participant E5}
  \label{fig:cactus}
    \centering
    \includegraphics[width = 0.9\columnwidth]{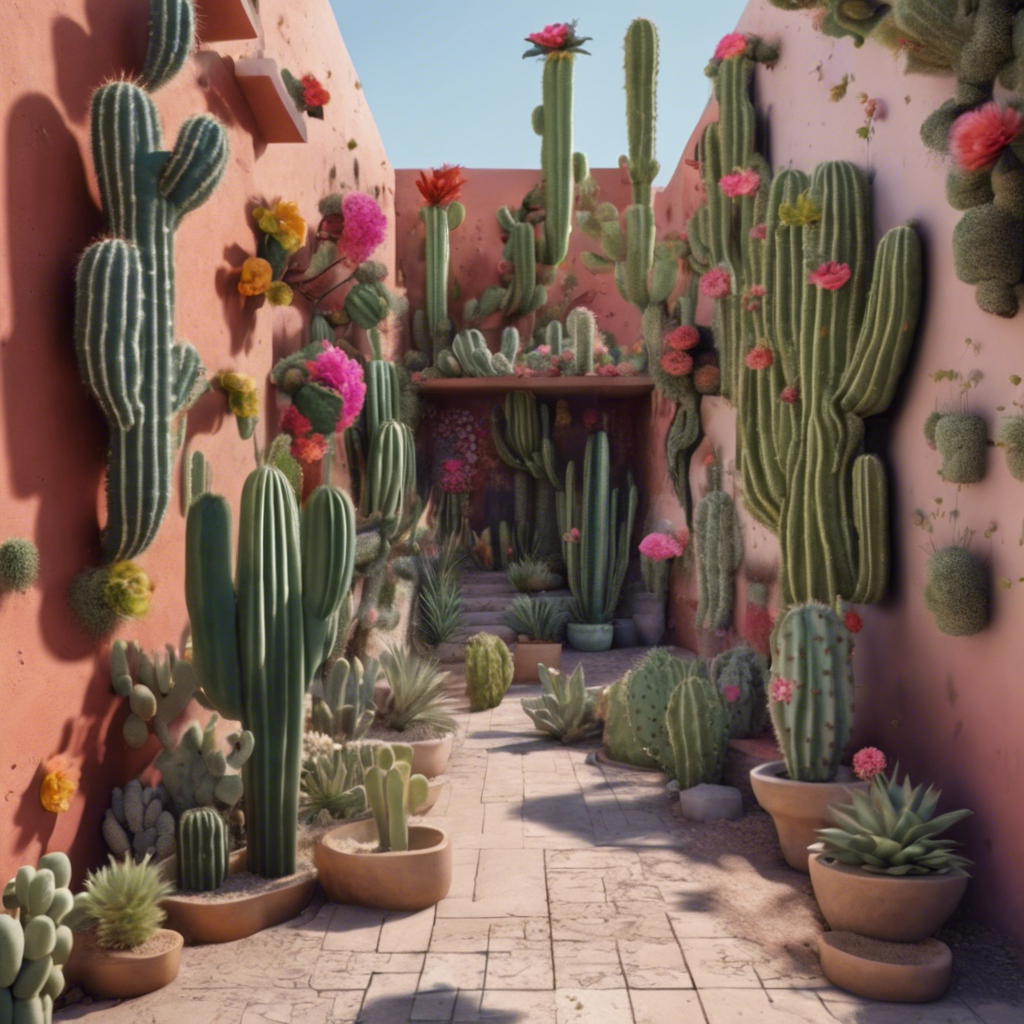}
    \caption*{\textbf{Prompt on Dream Studio:} Public space with novel, exciting ideas, roots growing on structures, cactuses are displayed, tiny little flowers Bougainvillea that grow around walls, trees, and flowers seem fun and look like they belong there.}
    \Description{Alley with small cacti on the floor and larger cactus emerging from the walls. Many cacti have orange and pink flowers.}
\end{subfigure}
 \caption{Images revealing hidden or unknown ideas}
\label{fig:sec411}
\end{figure*}

Conversational Imperfects, not necessarily aligned with traditional success metrics, revealed hidden or unknown ideas from participants. 
Through unexpected components and arrangements, images triggered memories, surprised the participants, and suggested follow-up questions to the facilitator. 
For example, an interviewee could not articulate a specific element through the interview that could recall her Mexican heritage until she saw a mural in one of the images, which reminded her of her mother's stories about her hometown. 
The picture made her recall significant childhood memories and find an element she would include in a public park to represent her cultural heritage (see figure~\ref{fig:mural}).

\begin{displayquote}[E2]
    \textit{I do remember my mom talking a lot about murals back where she's from, and then where I live too, there're a few. We have one of La Virgencita.}
\end{displayquote}

Similarly, another participant felt his Mexican heritage was unknown to him; images helped him to remember and discuss potential stereotypes about Mexican culture. 

\begin{displayquote}[E5]
    \textit{ I mean, I don't know anything because my dad was born here [In the US] (...). I just thought of cacti (...) and these beautiful flowers growing around my grandma's house where I lived in Highland Park. And I don't know what they're called. I think there's like a bee. And there's like these like tiny little flowers. And I think the leaves are the color of the flowers. The flowers are small. But they grow around like walls and stuff. And that's part of what I think of when I think about Mexican people (...). And they're just so beautiful. Every time you see them, they come in red, pink, and purple and grow close together.}
\end{displayquote}

When discussing figure~\ref{fig:cactus}, the interviewer and participant figured out the flower (i.e., Bougainvilleas) despite the image being inaccurate. 
From there, the participant shared how natural elements helped constitute his ideas about Mexico and how much he liked them in public spaces.

\begin{displayquote}[E5]
    \textit{Wow! That is cool. Yeah, just like that, a lot of natural stuff like trees and flowers. I love flowers on buildings and stuff like that (...). There are so many cacti around, but I think it looks good (...); I don't know much about Mexico. But I feel like this reminds me of my idea of it.}
\end{displayquote}

Discussing his Mexican heritage and working on the image helped the participant surface memories and potential stereotypes. 
The flowers and the cacti were the natural elements that came to mind when the participant thought about Mexico. 
These could come from popular media and stereotypes, like a description in which he mentions the Disney movie Encanto\footnote{Here there is a stereotype about Latin America since the movie is inspired in Colombia, not Mexico}: \blockquote[E5]{\textit{I'm thinking of the movie Encanto and the cacti and those beautiful flowers that that one girl makes}}. 
But they also come from memories and descriptions from his relatives: \blockquote[E5]{\textit{I mean, I don't know, it's interesting, but I've never been to Mexico. I just hear stuff that my grandma says}}.

In this case, Conversational Imperfects helped the participant and the interviewers discuss nuances about memories, stereotypes, and salient elements regarding cultural heritage. 
Despite the image being unrealistic, failing to depict accurate Bougainvilleas, and falling into stereotypes (e.g., cacti), it promoted a conversation about memories and natural elements, which constitute part of what builds the notion of Mexico in the participant. 
One practitioner reflected on IGAI during the workshop:

\begin{displayquote}[F5]
    \textit{This allows us to know the nuances of landscape elements important to the experience of space. Maybe they are particularly shaped by their cultural values, upbringing, or migrant identity. This allows richer conversations and their voices to be heard.}
\end{displayquote}

\subsubsection{Inspiring new ideas} 
\begin{figure}[h!]
  \caption{Park with elements made from Fungi discussed with participant E5}
  \label{fig:fungus}
    \centering
    \includegraphics[width = 0.5\columnwidth]{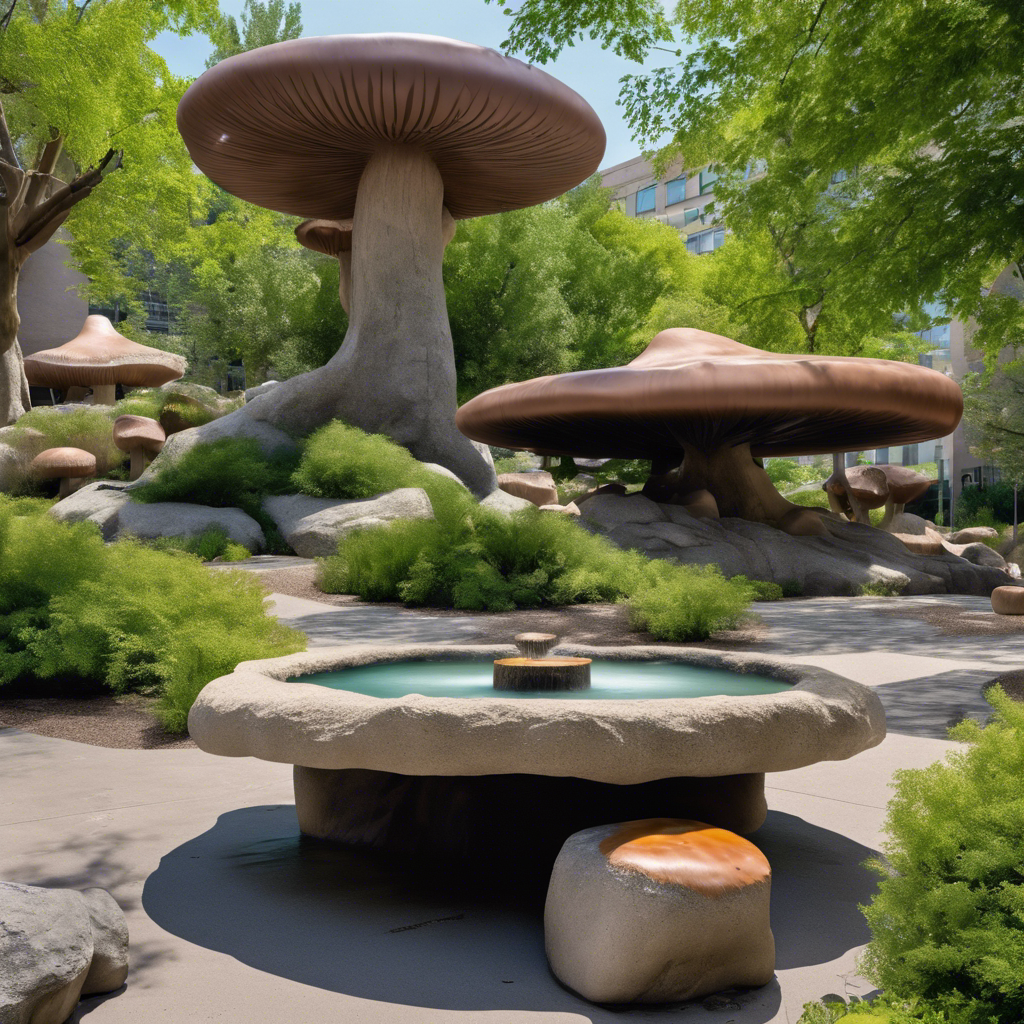}
    \caption*{\textbf{Prompt on Dream Studio:} It is a public space that feels wild and open and big but has small spaces; it is accessible,  has small, clean picnic areas, a medium-sized water feature with a fountain in the middle, fake rocks, and a fake mushroom that look natural and whimsical, and as they belong to the space, there are more trees.}
    \Description{A park with a fountain-like table with a seat that looks like the portion of a fungus. There are several large mushrooms to provide shade alongside trees.}
\end{figure}

Conversational Imperfects in the form of unrealistic and unexpected results inspired new ideas, facilitating discussions that moved away from common places.
For example, one participant expressed he liked fungi, and the IGAI generated a fantastical park where all elements were based on them (see figure~\ref{fig:fungus}). 
This image prompted ideas about designing using natural elements, which can lead to more sustainable spaces. 
In the practitioners' words, using IGAI \blockquote[F3]{\textit{has the potential to prompt and facilitate imaginative outputs}}. 
A researcher had similar reflections: \blockquote[F10]{\textit{Whimsical unrealistic images can be generative in helping participants to think more creatively}} and a practitioner added \blockquote[F2]{\textit{Even absurd images can be analyzed through a spatial lens to enrich design thinking for transformation beyond banal reproductions of what is already known}}.

\subsection{Finding 2: IGAI promoted space-aware conversations}
Conversations using IGAI as a boundary object promoted space-aware descriptions and experiential dynamics. 
Compared with the interviews conducted without IGAI, we observed that participants improved their descriptions of the space and their position within it.
Participants suggested that \textbf{articulating and crystalizing ideas through IGAI is particularly helpful for public space design, allowing them to be more concrete.}
First, participants tended to elaborate on more information about the spaces and their constraints when the generative outputs missed critical details and became Conversational Imperfects.
Second, echoing prior public space design literature, participants rarely came to clear, actionable design ideas in the first place; however, as they talked about the spaces, they deliberated more about the hidden design constraints and the relationships between people and the environments.
Finally, visual outputs supported more effective communication of these complex concepts. 
As such, we saw participants adopting more precise language and referencing the images to describe ideas for design.
In the following, we elaborated further on how working with IGAI enabled participants to articulate their ideas.

\subsubsection{Enhancing spatial and experiential descriptions}
\begin{figure*}[h!]
\begin{subfigure}[t]{0.45\textwidth}
     \caption{First iteration of the park with basketball court described by participant E3}
  \label{fig:spaceaware1}
    \centering
    \includegraphics[width = 0.8\columnwidth]{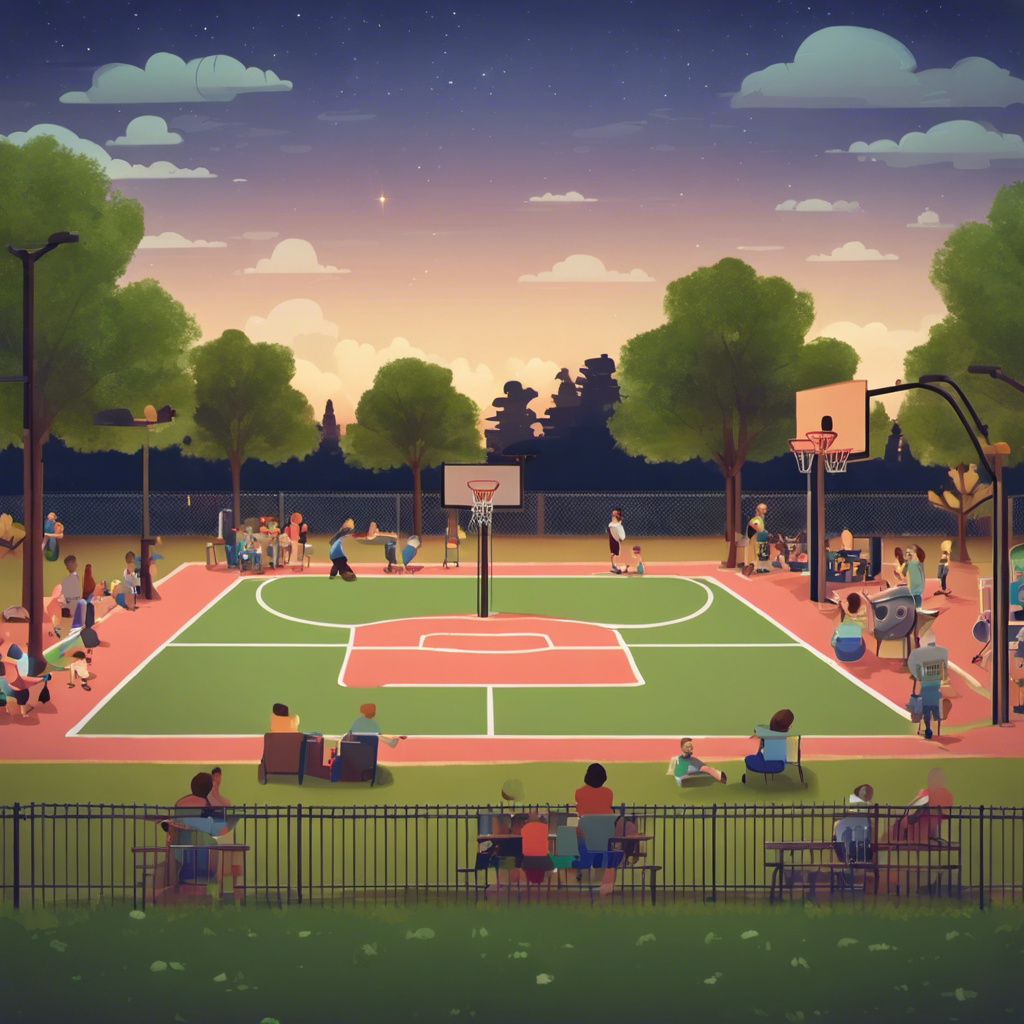}
    \caption*{\textbf{Prompt on Dream Studio:} Summer movie night at the park with live music with lights turned on, popcorn, activities, and games. There is a basketball court and two baseball courts. There is a playground. The park has cement, grass, and sand. The playground would be on the far right, there is a soccer field in the front, and in the center there are two baseball fields. }
    \Description{A park with a mixed basketball, soccer, and baseball field in the middle. There are a lot of people hanging out around the fields. There are trees and grass around the courts.}
\end{subfigure}
\hfill
\begin{subfigure}[t]{0.45\textwidth}
    \caption{Second iteration with participant E3 incorporating her position in the space}
  \label{fig:spaceaware2}
    \centering
    \includegraphics[width = 0.8\columnwidth]{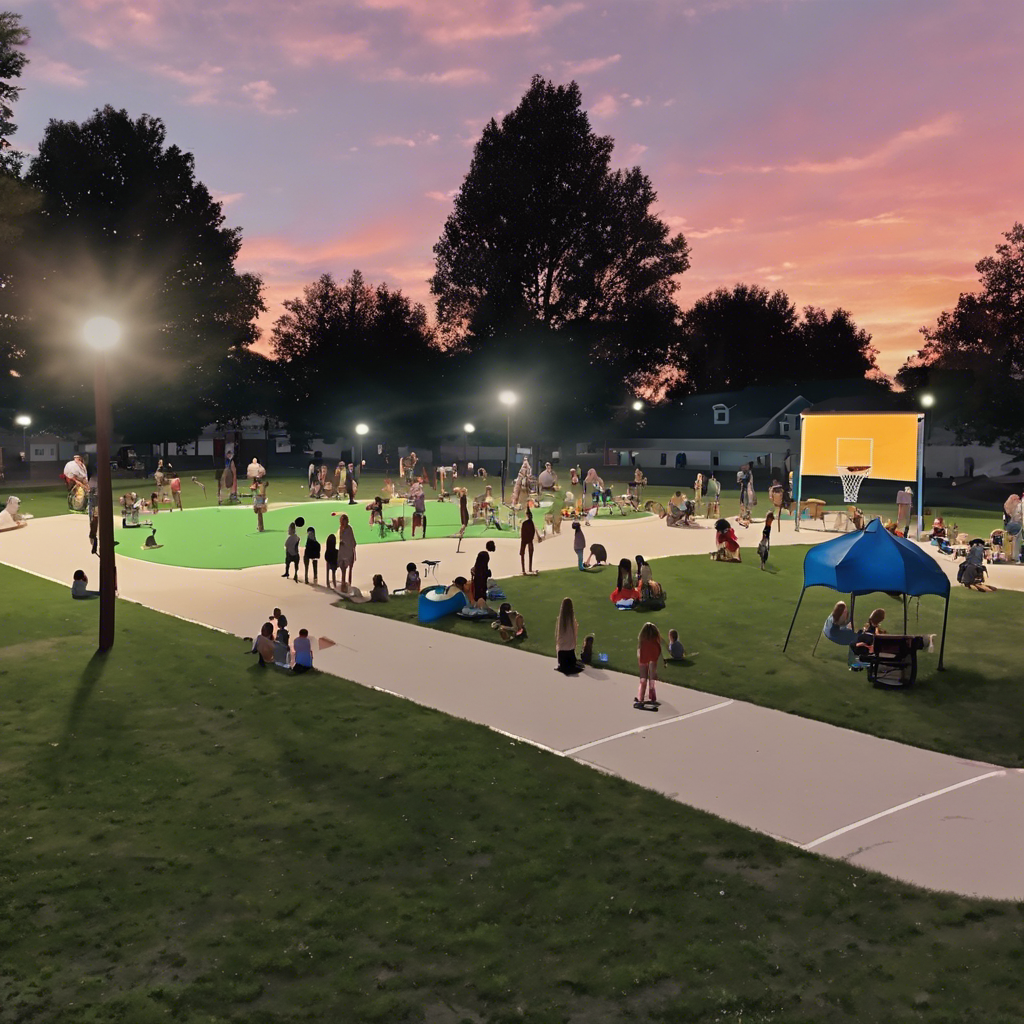}
    \caption*{\textbf{Prompt on Dream Studio:} Summer movie night at the park with live music with lights turned on, popcorn, activities, and games. There is a playground next to the basketball court. The park has cement, grass, and sand. The playground would be on the far right, there is a soccer field in the front, and in the center there are two baseball fields. There is a swimming pool.}
    
    \Description{A park with a mixed sports field in the middle. There are a lot of people hanging out. The point of view is from a corner, which is what the participant described.}
\end{subfigure}
 \caption{Iterations from participant E3 memory in a public space}
\label{fig:partE3}
\end{figure*}

Having an image to rely on helped participants improve their descriptions of spaces and experiences. 
As practitioners and researchers stated: "\textit{Images seemed to help participants think about colors of the space as a whole (e.g., shifting from flowers to built elements)}" (F10), "\textit{Images helped interviewees express an emotionally meaningful memory through color and texture}" (F6) and "\textit{Images promoted a more holistic thinking about the space}" (F8). 
When participants looked at the images, they could point out specific elements and describe their requirements and experiences with specific reference points. 
For instance, one participant described her memory as follows (see figure \ref{fig:spaceaware1}): \blockquote[E3]{\textit{It's been a while since I last went. It was nice. They had an inside and outside basketball court. And then they had, I think, two baseball courts. It was a pretty big park}}.

The image was far from accurate and contained a lot of imperfections, such as the mixed sports fields. 
The errors facilitated the discussion, prompting follow-up questions and more specific descriptions by looking at the image and providing specific references to improve it: 
\blockquote[E3]{\textit{The playground would be far right. And then, in front of the playground, there was a soccer field. And then towards the end, there was a basketball court, and then in the middle, there was a field, so that's where they played baseball, and then more towards the left, there was another big field (...) There should be trees, but it's like that [pointing to the image], far back.}}
Then, she incorporated information about her position in the space and the atmosphere. 
Although the final result is still inaccurate (see figure \ref{fig:spaceaware2}), she could point us to where in the space she used to hang out and the kind of atmosphere and set-up she liked. 
The participant described her position, pointing to different sections in the image.
\blockquote[E3]{\textit{We're here, instead of here. This is my point of view if I were to be in the playground, and this would be my aunt’s point of view. The soccer field would have been here, and then the basketball court would have been over here}}.

The latter description and interaction, prompted by Conversational Imperfects, contrasts with 1-shot descriptions during the interviews in 2023 in which participants described more activities rather than details about their position in the space: 

\begin{displayquote}[A2]
    \textit{It was like an outdoor stage. So, they obviously can host bands or entertainment. There are lots of trees. There's an area for the children, like swings. There are also large grass areas where you can play sports, so when I go, there are, you know, I revisit my old students who are now going to college or high school, and they like to give me an update. So we'll set up volleyball games, and we'll sit. We'll use the benches to sit down and celebrate people's birthdays that have passed, so we'll cut a cake and do that. Catch up with each other over food.}
\end{displayquote}

\subsubsection{Promoting experiential dynamics}
Using IGAI during the interview promoted a more active engagement since iterations over the images required interaction. As one practitioner reflected:
\begin{displayquote}[F3]
    \textit{[Using IGAI] introduced interactive and iterative elements to engagement, allowing the creation of experiential dynamics and reactions in real-time. These more interactive direct dialogues and activities help form bonds with people more authentically, promoting one-to-one interactions.}
\end{displayquote}

Moreover, by having the participant instructing the facilitator in real-time \blockquote[F8]{\textit{engagement with the participants increased since they could feel ownership of what we were thinking and designing}}.  
For example, one participant was impressed about how IGAI could quickly generate what they were describing: \blockquote[E7]{\textit{It was cool to see how everything I say can pop up like that. I don't know. It's really cool}}.

The dynamic of using IGAI helped trust-building and leveled interviewers and interviewees. 
As one of the practitioners reported:
\blockquote[F2]{\textit{Image creation helps dissipate tension or nervousness; it shifts the focus from a binary dynamic where the person asking questions becomes less prominent}}.
Here, Conversational Imperfects also played a role, since \blockquote[F8]{\textit{imperfect images could make the interviewer nervous, yet break the ice and allow for small talk and a better atmosphere}}. 
It was easier for participants to discuss elements and relationships in images they saw and could easily modify than describing spaces with no reference or with images that cannot be edited: \blockquote[F5]{\textit{Images got interviewees interested in the scene, and their ability to change the design became more familiar during the process. It helped the interviewees to engage with the questions and reduced resistance}}.

\subsection{Finding 3: Generative results are subject to facilitators' ability to manage reactions and interactions with the IGAI}

\subsubsection{The ability to navigate the unexpected}
\begin{figure*}[h!]
\begin{subfigure}[t]{0.45\textwidth}
    \caption{Image with patches of color in the sky described by participant E3}
  \label{fig:colors}
    \centering
    \includegraphics[width = 0.9\columnwidth]{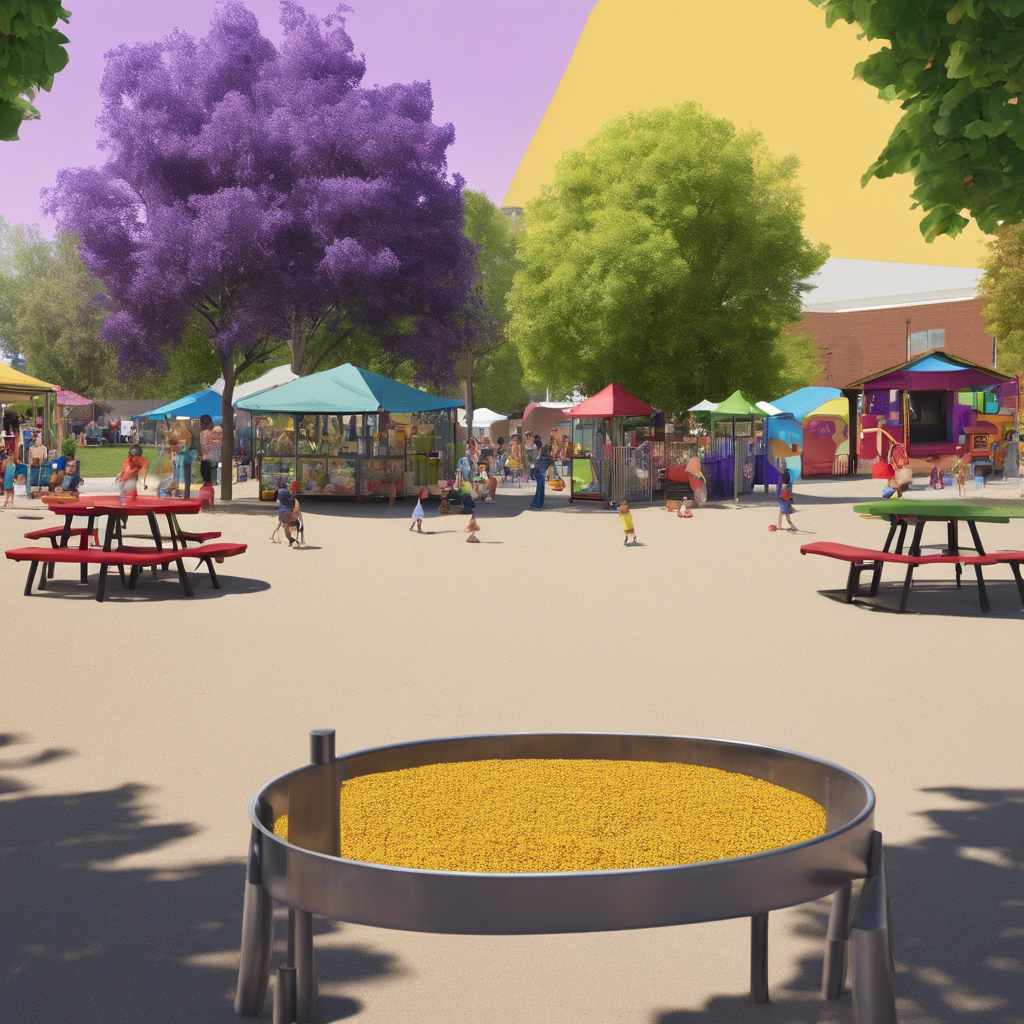}
    \caption*{\textbf{Prompt on Dream Studio:} A farmers market at the park with a playground with barbecues next to a playground. The playground has sand. There are also yellow, purple, and red flowers. A mural with different types of cultures as part of the community. Animals. Bathrooms close to the playground. There should be lights that are bright at night. There are food trucks in front of the park. There is also a parking lot.}
    \Description{Open square with vendors and round picnic tables. There is a container with sand. In the background, there are big trees. There are two patches of color in the sky, one purple to the left and one yellow to the right.}
\end{subfigure}
\hfill
\begin{subfigure}[t]{0.45\textwidth}
      \caption{River Garden Park improved by participant E3 using colors from the previous image}
  \label{fig:colors3}
    \centering
    \includegraphics[width = 0.9\columnwidth]{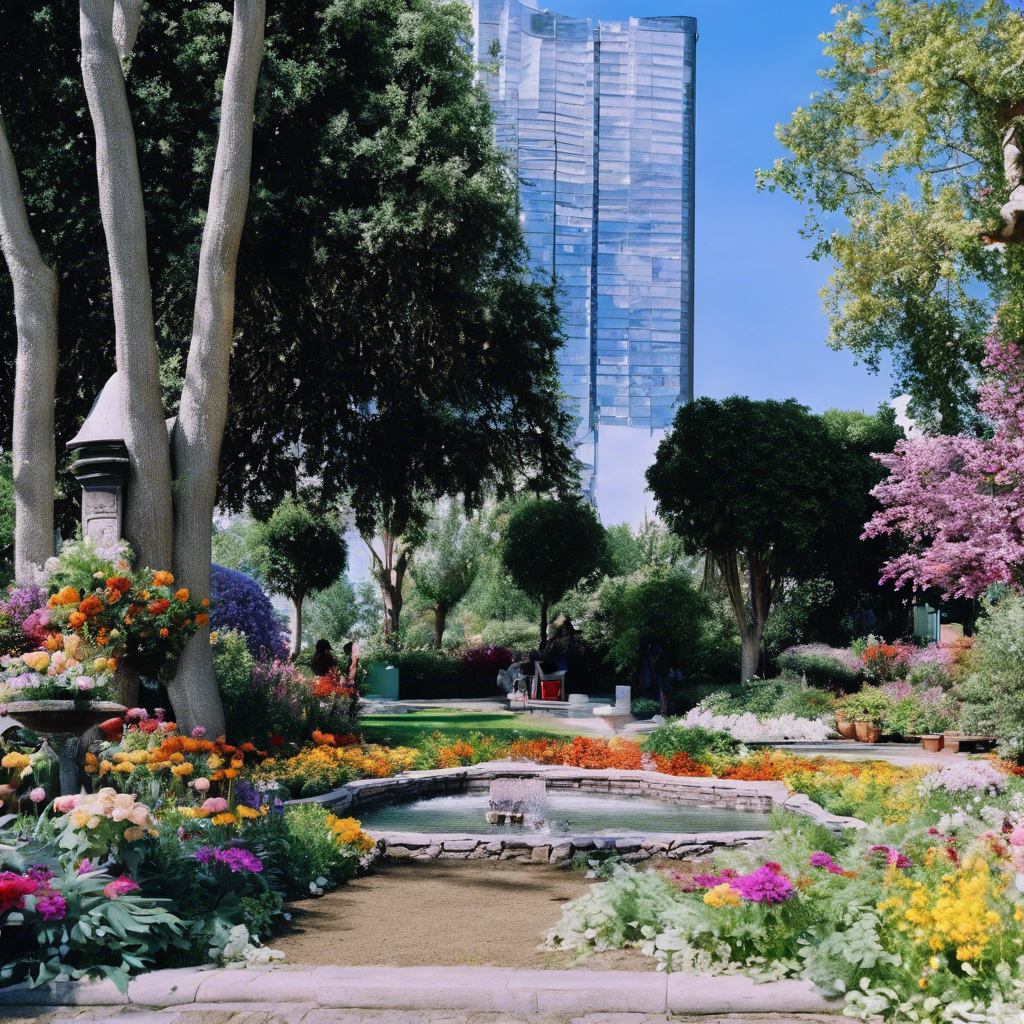}
    \caption*{\textbf{Prompt on Dream Studio: }Flower garden with bright colors and a water fountain. Shade from the trees and people resting under the shade. \\
    \textbf{Reference image strength\footnote{Controls the amount of noise added to the input image. Higher values allow more variations, but the output might not be semantically consistent with the input~\cite{esser_stable_2022}}: }35\%}
    \Description{River Garden Park improved with pink, purple, orange, and yellow flowers, a water fountain, and trees.}
\end{subfigure}
\caption{Images from phases 2 and 3 of participant E3}
\label{fig:partE3}
\end{figure*}

To extract valuable insights from images, facilitators need the ability to navigate unexpected results in a generative way. 
When interacting with IGAI and participants in real time, facilitators must be prepared to react to the unexpected, quickly reflect on it, and elicit new information to iterate or move on to the next stage without losing the connection with the participant.
One practitioner commented: \blockquote[F1]{\textit{Reaching mundane design outcomes and not knowing what to do with them can close the conversations}}.
For instance, participant E3 only identified the color palette she liked when she saw Figure~\ref{fig:colors}, which was then used to ask more specific questions about how colors would be present in improving the River Garden Park, resulting in a more detailed and realistic design as depicted in Figure~\ref{fig:colors3} which was selected based on the conversation about colors while scanning different options. 
However, to make that possible, the facilitator had to ask why that image, which was far from reality, was her favorite.

\subsubsection{Accurate images can close conversations}
\begin{figure}[h!]
  \caption{Memory in a public space described by participant E6}
  \label{fig:fakep0}
    \centering
    \includegraphics[width = 0.5\columnwidth]{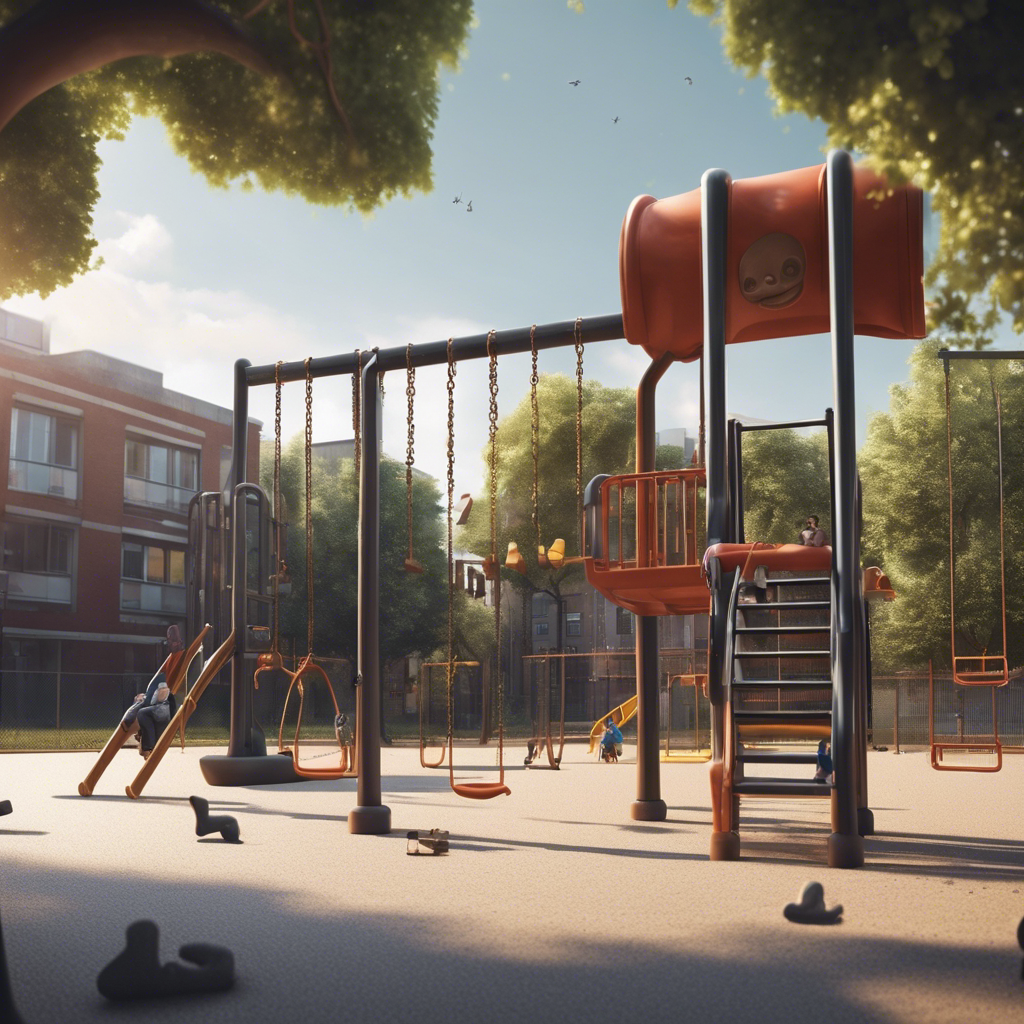}
    \caption*{\textbf{Prompt on Dream Studio: }Public space where people are in swings. There is a field next to them. In the playground, there are slides and monkey bars. It is the afternoon.}
    \Description{Public park with swings and games on the sand. There are some trees and a building in the background.}
\end{figure}

Fake Perfects (images aligned with the prompt and the interviewee's expectations) could close the conversations.
When the interviewee felt immediately satisfied with the results and the facilitators could not foster further exploration, the conversation was less rich in eliciting hidden values and desires or promoting new ideas.
Practitioners reflected: \blockquote[F5]{\textit{There were instances the interviewees did not comment on images to make further changes after only a few iterations, which did not reveal much information}}. 
The following conversation with Participant E6 provides an example when discussing figure~\ref{fig:fakep0}.

\begin{displayquote}
    \textit{\textbf{Facilitator}: Okay. Let's see what [prompter] has been creating [...]. \\
    \textbf{E6}: It was exactly like this one [pointing at one image].\\
    \textbf{Prompter:} Do you have any comments? Is there something that you'd like to add or something that it didn't capture?\\
    \textbf{E6:} No, I don’t think so; it's nice. Just the sand.\\
    \textbf{Prompter:} What about the sand?\\
    \textbf{E6:} It was more yellowish.}
\end{displayquote}.

\begin{figure}[h!]
  \caption{Image of a public square described by participant D1, which was accurate but failed to capture her cultural heritage}
  \label{fig:fakeperfect}
    \centering
    \includegraphics[width = 0.5\columnwidth]{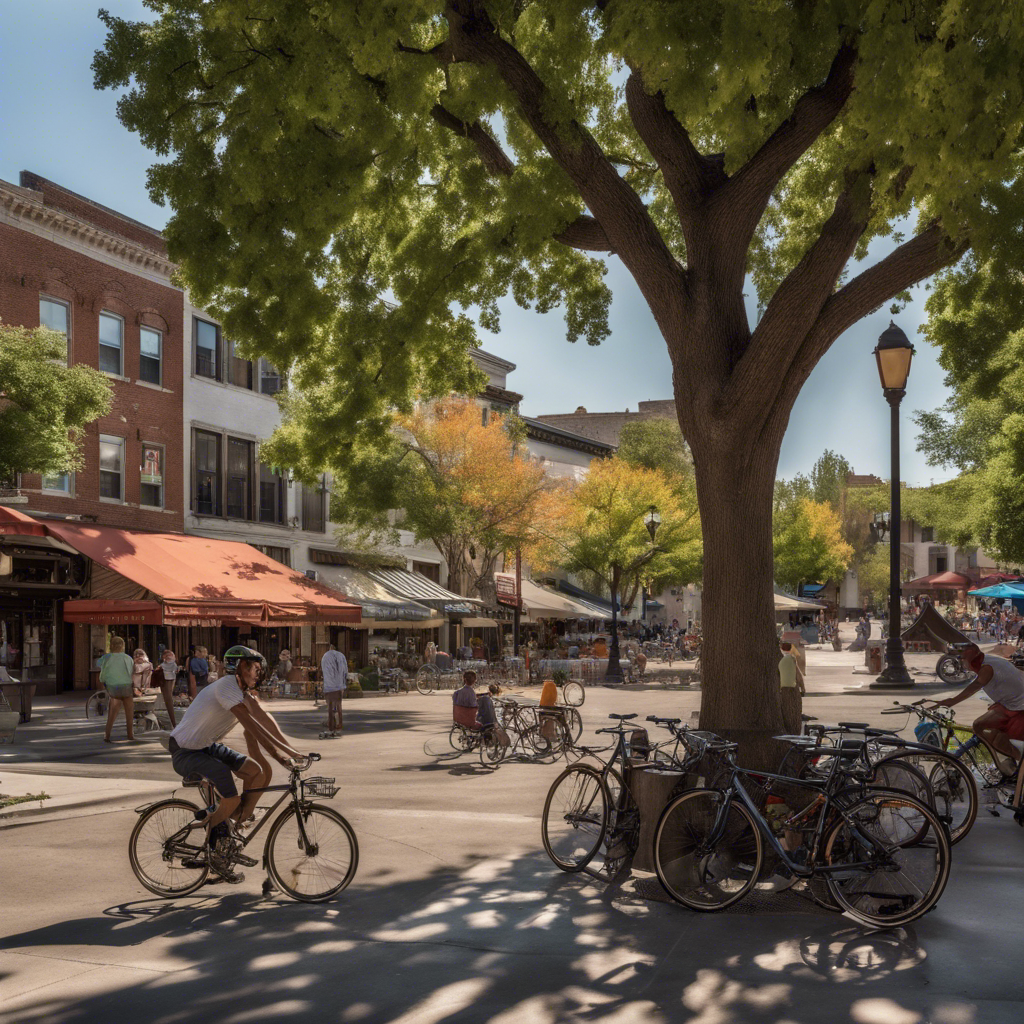}
    \caption*{\textbf{Prompt on Dream Studio: }In the heart of the neighborhood stands a huge square anchored by a majestic sycamore tree and a tranquil fountain. Cyclists weave through pathways, elders find shade, and children play football. Surrounding the square are inviting restaurants and cozy coffee shops, drawing the entire community together. As evening descends, people gather around the coffee shops beneath the comforting presence of the 100-year-old sycamore amidst modern office buildings.}
    \Description{Public square with a huge tree and bikes resting on it. There are old buildings and people walking and biking.}
\end{figure}

Moreover, Fake Perfects that were accurate in terms of the elements and relationships but misaligned with the participants' expectations could also fail to elicit valuable information. 
Practitioners reflected that some images could produce \blockquote[F2]{\textit{resistance or alienation from the participant when resonance between the interviewee and the image did not happen}}.
For example, during the piloting interviews, one of the images accurately depicted the described space, but despite having the right elements and relationships, it did not adequately represent the interviewee's cultural background and expectations (see figure~\ref{fig:fakeperfect}). 

\subsubsection{Understanding the human-AI interaction}
Practitioners and researchers reflected on the need to understand better how to prompt, how IGAI operates, and the potential biases of the system.
During the workshops, we discussed questions such as 
\blockquote[F8]{\textit{Which words are more influential in a prompt? Is AI picking different elements depending on the language? How is the complex interaction between biases of the AI and biases/objectives of the interviewer?}}, and 
\blockquote[F9]{\textit{Can the generated image create path dependencies?}}.

Prompting ability could influence how the conversations develop. 
Although, in our case, prompters generated images to elicit information and engage with the participants, this required practice sessions and pilot interviews within the team. 
One researcher reflected that 
\blockquote[F8]{\textit{protocols could help to improve the process with suggestions on how to prompt}}.
During the first round of interviews, failed prompts put pressure on some prompters, who became nervous when the images were too far from what was expected. 
Throughout the day, nervousness decreased from watching the positive reactions of the participants, but it might not have been the case.
Moreover, practitioners reflected that professional background influences the wording, which could lead to different outcomes. 
All prompters had the same background, but during the workshop, we reflected on how domain-specific words could influence the images. 

Practitioners wondered how initial images could anchor expectations and shape the session's development. 
There could be some path dependency in which the initial set of images or the reference image limits possible options regarding discussion and iterations. 
One practitioner reflected:
\blockquote[F1]{\textit{The use of partial arrangements on a chosen reference image perhaps limited the possibilities to unveil imaginaries; images showed some common sense but accepted and normalized an existing situation. For example, it was hard to go beyond limited expressions of translocality (e.g., colors, flags) after the first results}}.

During the workshop, practitioners worried about not knowing how much the AI was driving the interaction: 
\blockquote[F2]{\textit{We don't know if the AI is driving too much (...) there is obscurity in where the images are selected from and how they build upon}}. 
During the workshop, we discussed how, in this case, four agents' subjectivities (the facilitator \& the prompter, the AI, and the participant) interacted when creating the image, which could make the explainability challenge harder.
Relatedly, participants were worried about biases in both the AI and the prompter and how that could influence the session's path. 
For example, some generated images seemed idealized and US-looking or held stereotypes from specific cultures. 
During the workshop, we reflected:
\blockquote[F8]{\textit{Images without a starting point seem idealized and US-looking. To what extent can it work as an artifact to represent memories for migrants?}}.
We further discussed the complexities of representing the diversity of migrants' experiences and memories in images, especially those that seemed to generate American aesthetics inherently.

\section{Discussion}
IGAI offers interesting opportunities to promote a richer interaction between designers and stakeholders when designing public spaces. 
However, using IGAI as a boundary object to facilitate conversations between stakeholders and designers remains underexplored. 
Our study draws insights from using IGAI in a real-world participatory process. 
Based on our experience using IGAI as a boundary object, we defined Fake Perfects and Conversational Imperfects and analyzed how they influenced conversations in a co-design exercise mediated by IGAI.
We found a disconnection in how the success of generative output is evaluated by IGAI designers versus public space designers. 
Instead of focusing on realistic images that perfectly align with the text prompts, practitioners highlighted that generative images saw the greatest potential to prompt richer, more participatory conversations \textit{when they were imperfect}. 
Conversations mediated by IGAI became more space-aware and revealed hidden values and ideas. 
However, we also found that the generative results of using IGAI depended on how facilitators managed the experience.

Below, we reflect on how our findings point to further research on participatory methods using IGAI for public spaces. 
Moreover, we discuss how the HCI and CSCW communities could extend their research regarding IGAI from the designer-AI interaction dyad to HAII in a multistakeholder situation in real-world applications.

\subsection{IGAI as Design Support: From Object- to Process-Oriented Methods}
Our findings suggest a shift from object-oriented to process-oriented methods for the participatory design of public spaces. 
We reflect on three elements we distill from our findings: success metrics, the synchronicity of design narratives, and the interactive dynamics between practitioners, IGAI, and stakeholders. 
All three elements suggest that using the IGAI centers the experience in the conversation rather than the object itself. 
It is less important to achieve a high-quality sketch or a comprehensive list of features than to enable rich conversations that elicit values and promote stakeholder engagement.
We discuss research opportunities stemming from this shift.

\subsubsection{Towards contextual-success metrics}
How designers stir discussions becomes more important than generating the 'right' images. 
Our findings suggest that, when using IGAI as a boundary object in participatory processes, the goal is to promote conversation and leverage asymmetries between designers and stakeholders. 
Prior work suggested IGAI could bring different perspectives and promote creative approaches for designers \cite{chiou_designing_2023,shi_understanding_2023,wan_gancollage_2023}.
However, these results are encapsulated in the designer-AI interaction, so our work adds a new dimension. 
Our findings highlight the importance of deliberation around the prompts and results since IGAI might not foster creativity alone but through generative conversations.

Having metrics that adapt to different contexts can help calibrate methods and guidelines to promote a better HAII.
Our study extends prior work on metrics of success for IGAI that has focused on realism and text-image alignment~\cite{zhang_text--image_2023}.
Success metrics of IGAI-mediated participatory sessions could be expanded to how the generated images foster conversations and enable the designer to elicit deeper insights.
Incorporating ways of measuring Conversational Imperfects as success would enrich how IGAI can be used for participatory processes.
Proposed metrics should be able to capture success in the relationship between actors and the IGAI in the context in which it is being deployed.
However, while our findings highlight the value of IGAI in promoting meaningful conversations and engagement more than the realism and text-image alignment, the latter can still play a significant role in the process.
Having IGAI generate successfully matched imagery that can support the process of obtaining participant feedback is still of great interest from the designer’s perspective, largely because it can potentially empower participants’ sense of total contribution during the participatory process.
In that sense, providing designers easy-to-understand ways of setting the system's level of randomness and the text-image alignment could help them manage how they intentionally generate imperfect images to promote conversation.

Future research should explore ways of measuring success that match the needs of designers using IGAI as a boundary object in participatory design. 
We suggest analyzing the ability of IGAI-generated images to (1) foster robust discussion between designers and participants, (2) enable the designer to elicit deeper insights, and (3) reflect a qualitative impression of the participants’ psyche that can both empower them alongside providing accurate sketches to designers.
Moreover, our results suggest that there is value to the randomness and inaccuracies that currently exist in IGAI. 
As the technology continues to evolve, developers should consider preserving the ability to introduce randomness in the process. This could be achieved by providing users with easy-to-understand settings to control the level of randomness.

\subsubsection{Design narratives' synchronicity}
Using IGAI changes the synchronicity of design narratives when using images by introducing real-time image co-design. 
Designers often use reference images, such as photographs, collages, illustrations, and computer-generated images (CGI), as boundary objects to facilitate collaboration with participants. 
However, these images are typically asynchronous, meaning they must be created before the participatory experience, allowing for limited or no iterations based on the participants' input.
For instance, CGI often depends on pre-existing templates, incurs high costs, and lacks the adaptability to reflect specific communities' unique aspirations and constraints. 
Consequently, the participatory conversation evolves \textbf{from} these static images, constraining the dynamic exchange of ideas~\cite{star_institutional_1989}.

In contrast, IGAI enables synchronous design narratives by allowing images to be co-created with participants in real-time. 
This dynamic process ensures the design exercise evolves concurrently \textbf{with} the images, fostering a more iterative and responsive participatory experience~\cite{disalvo_design_2009}. 
The ability to generate and modify images on the fly means that participant feedback can be immediately incorporated, leading to a more relevant and inclusive design process. 
By facilitating real-time visualizations, IGAI helps bridge the gap between designers and community members, fostering how diverse perspectives are integrated into the design of public spaces~\cite{cantrell_responsive_2016}. 
This synchronous approach could enhance the legitimacy of the design outcomes and empower participants by making them active contributors to the creative process.

There are tradeoffs when deciding which boundary object to use when designing public spaces.
IGAI can provide a synchronous experience of co-design, in which designers can work with stakeholders to iterate over different images while discussing desires and preferences without too much expert knowledge and resource constraints.
However, when using IGAI, the designer loses control over prompting specific reactions that could be carefully tailored in a reference image. 
Furthermore, the designer's ability to prompt and understand how the IGAI works will mediate the process's success.
In contrast, reference images are usually incorporated in the early stages to express design intent or to explore different project types in a controlled and curated manner.
However, more control comes with restrictions, such as setting a stronger path dependency and being unable to react to the unexpected since reference images usually depict ideas already existing in the built context. 
For instance, participants may come from cultural backgrounds unfamiliar to the designer, which risks not having appropriate or relevant reference imagery on hand. 
Depending on the methods, generating reference images can have material restrictions such as cost, hardware requirements, and time.
Future research should compare methods and explore hybrid options to shed light on the specific advantages and disadvantages through different stages of the design process.

\subsubsection{Pointing elements, eliciting values}
Conventional participatory processes often focus on programmatic activities, uses, and objects (amenities). 
While these elements are crucial to the design and functionality of public spaces, a more profound approach involves eliciting and integrating the community's underlying values into the design process, not only highlighting generic demands on greenery, shade, and amenities. 
This transition from focusing on physical elements to eliciting values can significantly enrich the design of existing and new public spaces.

Integrating values into the design process involves translating these values into actionable (design) elements manifested in the physical landscape. 
This process requires interpreting expressed values in ways that can be integrated into the design process and practically implemented, such as creating communal seating areas to foster social connectivity, incorporating sustainable materials to reflect environmental values~\cite{disalvo_design_2009} or the ideas of transcultural murals, vegetation, and colors revealed in the interview process. 

Tools like IGAI enhance this process by providing dynamic visual representations of design scenarios. 
They were useful in prompting participants to express more of their sentiments and personal values. 
IGAI allows for an iterative design process in which participants can see and adjust visualizations in real-time, ensuring that the final design closely aligns with their aspirations and needs. 
This approach fosters a sense of ownership and connection among the designer and the stakeholders.
This transition acknowledges that the success of public space design is not solely dependent on the program and physical features but also on how well it aligns with the community's values and needs~\cite{sanders_co-creation_2008}, uncovering and reflecting communities' fundamental beliefs, aspirations, and priorities. 

\subsection{The Expert is No One: IGAI as a Medium for Public Engagement and Participatory Processes}
Findings from our study provide insights into how adopting emerging technologies influences participatory, co-design processes.
This section discusses how IGAI can more effectively engage multiple perspectives from stakeholders.

\subsubsection{Addressing knowledge-based power dynamics in participation}
Adopting IGAI tools to mediate the conversations between designers and stakeholders impacts the power dynamics in such interactions.
\textit{As neither designers nor the stakeholders are experts in genAI, working through IGAI tools allows them to explore design ideas in more equal positions.}
To compose effective prompts as IGAI inputs, both designers and stakeholders tended to articulate their design visions and surrounding environments with extensive details.
This helps filter out their assumptions about what each other knew (and didn't know) about one another.
From a collaboration standpoint, this practice establishes a clearer common ground.

Prior literature has raised numerous issues about having domain experts lead participatory processes~\cite{delgado_participatory_2023, Khan_smart_city_2017}.
While domain knowledge helps formulate effective conversations (e.g., asking participants questions that can directly lead to actionable design ideas), it can often lead to unintended power dynamics during participatory processes.
Findings from our study show that introducing IGAI to stakeholder-designer interactions can mitigate this problem and create more welcoming spaces for public engagement.

Moreover, participants face barriers in conventional workshop methods (e.g., posters, base maps, photo boards, surveys, etc.) since these mediums require participants to rely on their own skill sets, which can prove challenging, especially within time constraints. 
Participants struggle to convey their ideas through these means because they may not have experience translating their wishes into a designer’s feedback kit. 
When this is the case, participants immediately resort to writing down their thoughts or ideas for designers to make sense of later. 
While textual feedback is straightforward and useful, the IGAI imaging exercise can promote more creative and imaginative output from participants.

\subsubsection{Reflection of values}
As participants elaborate more on the public spaces and their design ideals, they simultaneously express more of their embedded values.
Indeed, each individual sees the same landscape differently; these lenses are often shaped through their personal, contextual, and social backgrounds~\cite{Boer_Fischer_2013}.
In public space design --as well as other forms of publicly engaged design-- extracting and comprehending participants' values is more important than getting them to visualize their ideas in perfect images.
Understanding stakeholders' values behind their desired design enables designers to take away more generalizable insights for future work.
In fact, some might argue that imagery properly capturing stakeholders' values might diverge from the ``real'', physical scenes~\cite{Boer_Fischer_2013}.

In this regard, IGAI's capabilities, which allow individuals to create content based on their visions, are particularly useful for extrapolating their values and perspectives.
One can also interpret the utility of applying IGAI in participatory public space design in an alternative fashion: 
\textit{Instead of relying on AI to comprehensively present details about the space, designers should turn to AI as a design probe that prompts stakeholders to deliberate and express their values related to the space.}
While we worked with our participants throughout the study, using IGAI to depict their views and desires for public spaces also effectively challenged \textit{global knowledge}.
For public space design, designers typically come to work in an environment that has existed for several years, decades, or even centuries.
The freedom to generate new visions enables participants to showcase the same landscape from their points of view --and allows designers to reimagine the design subject through different cultural and social perspectives.

\subsection{Open Research Questions for Leveraging Technologies to Facilitate Public Engagement and Participatory Processes}

Our study also reveals numerous open-ended questions for applying IGAI and other emerging technologies in public engagement and participatory processes.
Here, we organize them into themes and propose the following research agenda:

\subsubsection{Expertise, training, and new roles in facilitating participatory design}
As discussed in the previous section--besides their domain knowledge in public space design--, participants' expertise in IGAI (or other forms of emerging technologies used to mediate participatory processes) determines their experience during collaboration.
More experienced IGAI users could typically attain more accurate generative images, but these perfect outputs often failed to stir more engaging conversations.
Similarly, high expertise in using the IGAI but no domain knowledge could fail to elicit valuable information for design.
With these ``fake perfect" images, experienced IGAI users with no domain knowledge could also fail to elicit useful information for design.
On the other hand, domain experts without sufficient IGAI literacy could overlook biases and potentially harmful content in generative outputs.
Future research should explore the complex interplay between domain expertise and IGAI usage expertise, so that users at differing proficiency levels could all leverage IGAI tools effectively.

Moreover, while being IGAI novices allows all participants to engage in the design conversations more equally, we expect continual growth in the public's AI literacy, making studying differences in knowledge and skills even more relevant.
How will that change their experiences during IGAI-mediated co-design processes?
How will that influence the effectiveness of adopting GenAI and emergent technologies in public engagement workflows?
Moreover, should designers receive special training to apply technologies to mediate their participatory processes effectively?
Should there be new roles -- such as facilitators supporting designers and stakeholders through tech-mediated public engagement and participatory processes?
Together, we ask \textit{who and what types of new expertise should designers acquire if they are interested in incorporating novel technologies in their participatory workflows.}

Considering educational materials that help establish the aforementioned expertise, we distill three domains. 
Technical training should equip facilitators to craft effective prompts and iteratively refine outputs based on participant feedback. 
Conversely, qualitative skills are likewise important for uncovering community values and needs, ensuring that AI-generated images reflect local contexts, cultural sensitivities, and nuances of difference. 
Finally, ethical training should be promoted. 
It should address the responsible use of AI and emphasize fairness, transparency, and privacy considerations to mitigate biases and uphold ethical standards in processes using IGAI as boundary objects. 
Again, a comprehensive learning agenda is also an important research area.

The three domains above are discipline-agnostic and could be applied outside of designing public space.
Only one of the researchers and none of the research assistants were experts in facilitating public participation processes and qualitative methods.
The only training required for the research assistants was to familiarize themselves with the IGAI system's features and practice prompting for a short period of time.
However, the researchers had domain-specific knowledge about designing public spaces, which was relevant to direct the conversation toward what was relevant in the design process.
Future research should explore how different skill levels in the three mentioned domains, plus the specific domain of the participatory process being conducted, influence the value extracted from using IGAI as a boundary object.

\subsubsection{Biases in IGAI outputs.}
Our findings revealed that practitioners were worried about potential biases in the AI system and how that could harm participants both during the session and through the resulting design.
Prior work has shown biases in IGAI output, and certain communities are more susceptible to inferior experiences caused by such biased content~\cite{vazquez_taxonomy_2024,zhou_bias_2024,qadri_ais_2023}.
Most existing work focuses on biases related to human figures in generative output~\cite{vazquez_taxonomy_2024,zhou_bias_2024,jha_visage_2024}, while culture and spatial biases are less studied because they are harder to capture in explicit forms~\cite{qadri_ais_2023}.
Thus, to use IGAI as a boundary object in participatory processes, future research should explore how biases can affect participants across different communities and how such consequences can be mitigated by technical design and methodological guidelines~\cite[e.g., ][]{jha_visage_2024}.

Our findings suggest that imperfect images can promote conversation, collaboration, and, in turn, better design.
This finding could also be expanded to studying biased images. Namely, how might one utilize imperfect generative output to initiate discussions about biases, generate awareness among participants, and critique stereotypical narratives?
For example, biases in IGAI could be used to teach participants about identifying biases and social injustice~\cite{apiola_first_2024}
Using community-centered research to study AI-generated images~\cite{qadri_ais_2023} can reveal how to use biases in images as a generative element in conversations.
Future research can expand on the line of work and explore mitigative approaches to addressing biases during participatory processes.

\subsubsection{Scalability of technology-mediated participatory methods.}
While participants acknowledged the advantages of adopting IGAI during the work processes, they held differing opinions about its scalability.
Like other workstreams that advance through engaging with the public, public space design can benefit from accumulating perspectives from a wider range of stakeholder groups.
However, the current IGAI-mediated method only applies to conducting participatory sessions with limited participants.
\textit{Whether and how to scale this protocol -- or, whether there should be alternative IGAI-mediated protocols that allow for more participants' engagement at once} --- these remain open-ended questions that we encourage future research to investigate.

It is worth noting scaling up public engagement and participatory methods requires more than increasing the sample size of participants~\cite{Delgado_NeurIPS_2021,delgado_participatory_2023}.
Engaging with more participants yields more diverse perspectives for design, but it can also overshadow the value of human intervention~\cite{seaver2021care, lempert2016scale, tsing2012nonscalabilitythe}.
Synthesizing and applying these various insights then becomes a key challenge when it comes to scaling participation.
Here, we pinpoint a few dimensions for consideration:

\textbf{\textit{Simultaneity of sociotechnical dynamics.}}
As designers cultivate varying perspectives from participants, many of their considerations are rooted in sociotechnical dynamics that operate simultaneously in the same environment but at different magnitudes.
In public space design, one might hear concerns about micro-level interventions in specific neighborhoods, while others share macro-level analyses of city-wide policies, regulations, trends, and patterns.
Differentiating stakeholders' perspectives by their scopes and mapping out how these multiscalar dynamics take place in parallel can be a useful first step for designers to tackle participation at larger scales.

\textbf{\textit{Spatial (in)equality.}}
Synthesizing perspectives from different stakeholder groups, designers might also see conflicts emerging across different communities --as well as conflicts occurring at the local versus global levels~\cite{Jacobsen_Challenging_the_space_2018, Harvey_2013}.
Namely, ideal design solutions for a specific community might contradict those for a broader society.
In this regard, designers should balance interests across different groups while actively prioritizing the interests of historically underrepresented, marginalized communities.
Managing such tradeoffs can be a complicated task, while approaches from other disciplines (e.g., macroeconomics, sociology, government science) might offer helpful insights~\cite{Lean_impact_2019, creation_of_social_value_2016, callon_essay_1998}, they have not yet been widely endorsed by HCI researchers.
Here, we encourage the CSCW community to explore these multidisciplinary approaches.

\textbf{\textit{Interaction over time.}}
Participation processes with a larger group of stakeholders are often projects that take a lot of time and resources.
Practitioners should consider the temporal component to make a broader impact through participation.
Using public space design as an example, one should ask how urban development, cultural transformations, technological support, etc., might change over time as they extract insights from participatory sessions to develop the design process.
Moreover, future research should explore whether and how IGAI and other emerging technologies (e.g., LLM-based simulation) can facilitate participants' and practitioners' envisioning of both short—and long-term societal changes.

\textbf{\textit{Operational feasibility.}}
The usefulness of public engagement and participatory methods goes beyond satisfying all stakeholders' needs.
Oftentimes, the success of participation is resource-dependent.
Understanding financial, expertise, personnel, and technical requirements is critical through engaging with more participants.

\subsubsection{Integration with other design methods and tools.}
Using IGAI as a boundary object can complement existing design methods and reach broader, more diverse audiences.
IGAI could be deployed on other platforms (e.g., Messaging apps, online discussion forums) and integrated in different methodological protocols (e.g., interviews, focus groups).
For example, prior work~\cite{bowen_metro_2020,bowen_metro_2023} has leveraged imagery content to facilitate co-design sessions at both small and large scales.
Moreover, one can also adopt IGAI for speculative methods~\cite[e.g., ][]{wong_infrastructural_2020,johannessen_speculative_2019,clarke_more-than-human_2018,ciaramitaro_imagination_2024,blythe_research_2014,grand_design_2010}, such as using IGAI as provocative content in design fictions~\cite{ostvold_ek_speculative_2024,blythe_artificial_2023,friedrich_speculative_2024}.
We encourage future research on how to incorporate IGAI into existing co-design methods and the creation of new ones that enable a broader and more equal participation of the public.

\subsubsection{Guidelines, standards, and protocols}
Besides accounting for the above-mentioned considerations, designers must address numerous environmental and societal infrastructure constraints.
The practice of public space design—-as well as other forms of design that involve public engagement—-has a substantial impact on society as a whole.
As such, it is arguable whether additional protocols should be added to existing state and municipal ordinances to ensure benign outcomes of such design practices (and the use of novel technologies in such practices).
Furthermore, designers must also ensure that they adhere to local policies when conceiving their design solutions, and having official guides to help navigate these regulatory frameworks can be beneficial.
Constructing such guidelines requires expertise not only on the technical front but also in participatory methods and design with public engagement.
We encourage joint forces from multiple disciplines to tackle such challenges.

\section{Conclusion and Limitations}
In this paper, we present the findings stemming from using IGAI as a boundary object to design public spaces. 
We expand previous work by conceptualizing Fake Perfects and Conversational Imperfects to surface a disconnection between conventional metrics of successful technologies and more nuanced approaches in participatory processes.
Using IGAI in the participatory design of public spaces promoted a shift from accurate to imperfect images, which in turn promoted richer conversations. 
IGAI-generated images could reveal hidden or unknown ideas and promote space-aware conversations.
However, the success of IGAI as a boundary object between the designer and stakeholders is subject to the designers' capabilities in managing unexpected and surprising results. 
We show potential opportunities for using IGAI in participatory processes along with limitations, surfacing complex dynamics that require further exploration, which is discussed in a proposed future research agenda.
We extend the HCI literature on IGAI by moving away from studying the designer-AI dyad and putting IGAI as a boundary object for multiple parties in a real-world experience. 
Through this work, we expect to promote research and reflection in the HCI community about the nuances IGAI introduces when designing to solve wicked problems such as designing public spaces.

Our work has limitations that we expect future research to address. 
Our findings stem from working with a specific community in Los Angeles, US, with different immigrant sub-communities and are constrained by the context. 
Moreover, our research team had more limited engagement with community members than we would have liked. 
Because we had one lead researcher interviewing participants and a research assistant using the IGAI for each interview, we were limited to scaling up the total sample size of our participatory sessions.
Future research might consider collective ways to include more participants in the IGAI-mediated process.
We encourage the HCI and CSCW communities to continue exploring the potential of new technologies to support human-centered participation that engages with broader and more diverse stakeholders.

\begin{acks}
We acknowledge generous funding from the Migrations Initiative and the Office of the Vice Provost for International Affairs at Cornell University. We acknowledge our amazing research assistants who were key in deploying our research: Achilleas Souras, Alberto Salgado, Jordi Prieto, Yike Xu, and Xiaochang Qiu. We also acknowledge Studio MLA and Mujeres de la Tierra for their collaboration, support, and engagement in the project. The views expressed in this manuscript are our own, as are any errors or omissions.
\end{acks}

\bibliographystyle{ACM-Reference-Format}
\bibliography{references, refs}
\appendix

\section{Description of IGAI software used in the IGAI-mediated interviews}
\label{ap:software}
  \begin{tabularx}{0.85\columnwidth}{|p{0.15\columnwidth}|X|}
    \toprule
    Software&Description\\
    \midrule
    Dream Studio & Software based on the open-source Stable Diffusion model \cite{rombach_high-resolution_2022} developed by Stability AI. Users prompt the system in a free text box and can configure style through a drop-down menu. Additionally, users can use prompt weighting to define a hierarchy, the number of elements to incorporate (e.g., colors, objects), and negative prompting to instruct the model on what not to generate. Users can start from scratch or an existing image. There is further customization regarding image size, count, generation steps, prompt strength (i.e., how much the image will portray the prompt), and seed (i.e., initial noise) \cite{stability_ai_dreamstudio_2023, stability_ai_stable_2023}. \\
    \midrule
    DALL-E & It is a transformer language model developed by OpenAI that can receive text and images to generate images from scratch or regenerate regions of existing images \cite{openai_dalle_2021}. From its original version, DALL-E has been improved in terms of accuracy following the prompts, quality of the images, and safety \cite{openai_dalle_2022, betker_improving_2023}.\\
    \midrule
    Photoshop AI & The software works through the Adobe Firefly model developed for the Adobe suite apps and services. The Firefly models were trained on collections of Adobe stock images, licensed content, and content from the public domain to ensure Firefly does not generate images based on other people's brands and intellectual property. The software allows the user to generate images from scratch and to generatively edit sections or specific elements (e.g., background) of pictures \cite{adobe_photoshop_2024, hutson_expanding_2023}. \\
   
  \bottomrule
\end{tabularx}


   


\section{Images and Prompts from Stage 1}
\label{annex1}

\begin{figure}
     \centering
     \begin{subfigure}[h!]{0.78\textwidth}
         \centering
         \includegraphics[width=\textwidth]{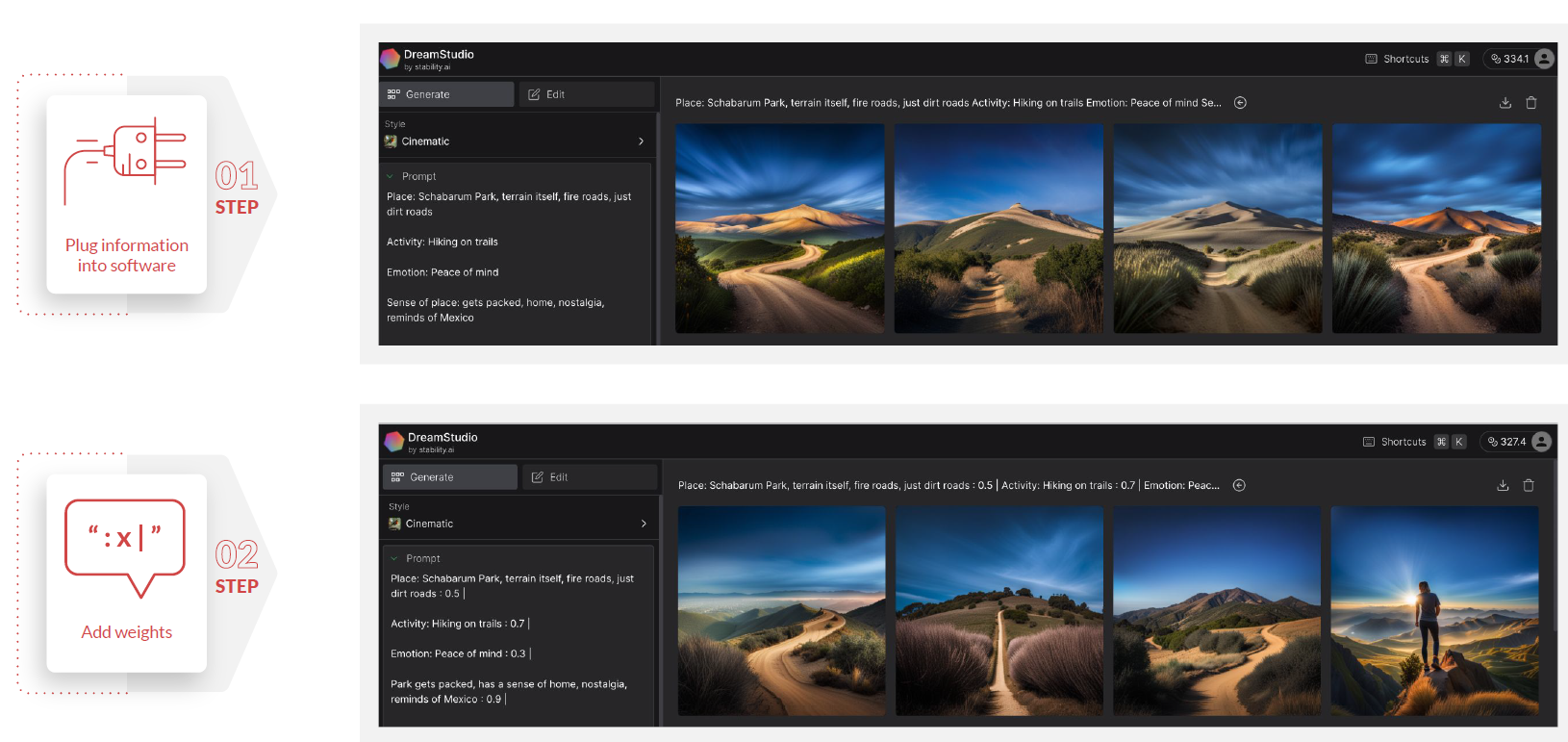}
         \caption{Stage 1 - Steps 1 and 2}
         \label{fig:Stage1(1)}
     \end{subfigure}
     \hfill
     \begin{subfigure}[h!]{0.78\textwidth}
         \centering
         \includegraphics[width=\textwidth]{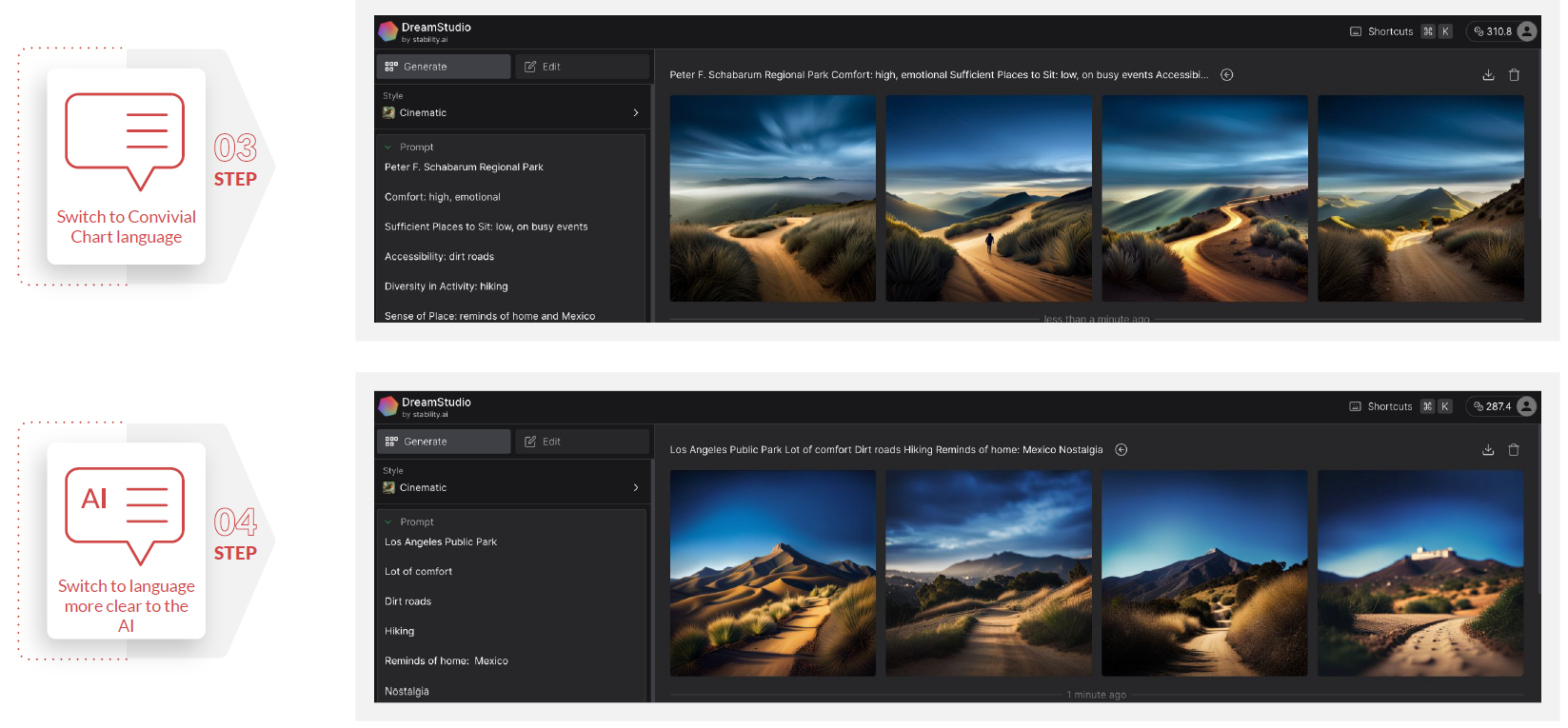}
         \caption{Stage 1 - Steps 3 and 4}
         \label{fig:Stage1(2)}
     \end{subfigure}
     \hfill
     \begin{subfigure}[h!]{0.78\textwidth}
         \centering
         \includegraphics[width=\textwidth]{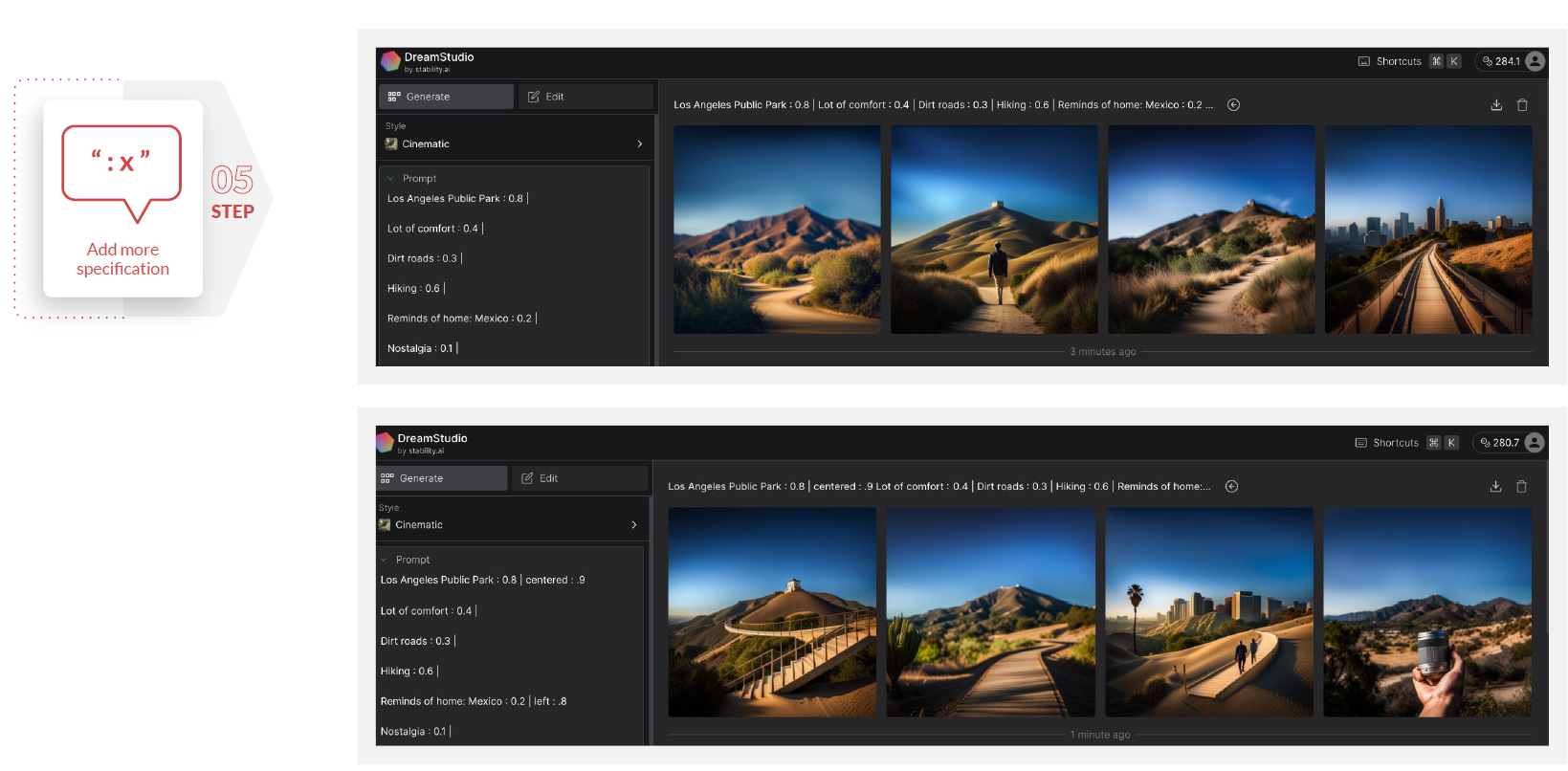}
         \caption{Stage 1 - Step 5}
         \label{fig:Stage1(3)}
     \end{subfigure}
        \caption{First Stage of Trials}
        \label{fig:FirstStage}
        \Description{Three figures show five steps to generate the image, the prompts, and the images generated through the process. The steps are (1) Plug the information into the software, (2) Add weights, (3) Switch to Convivial Chart Language, (4) Switch to language more clear to the AI, and (5) Add more specifications.}
\end{figure}

\begin{figure}
     \centering
     \begin{subfigure}[h!]{0.95\textwidth}
         \centering
         \includegraphics[width=\textwidth]{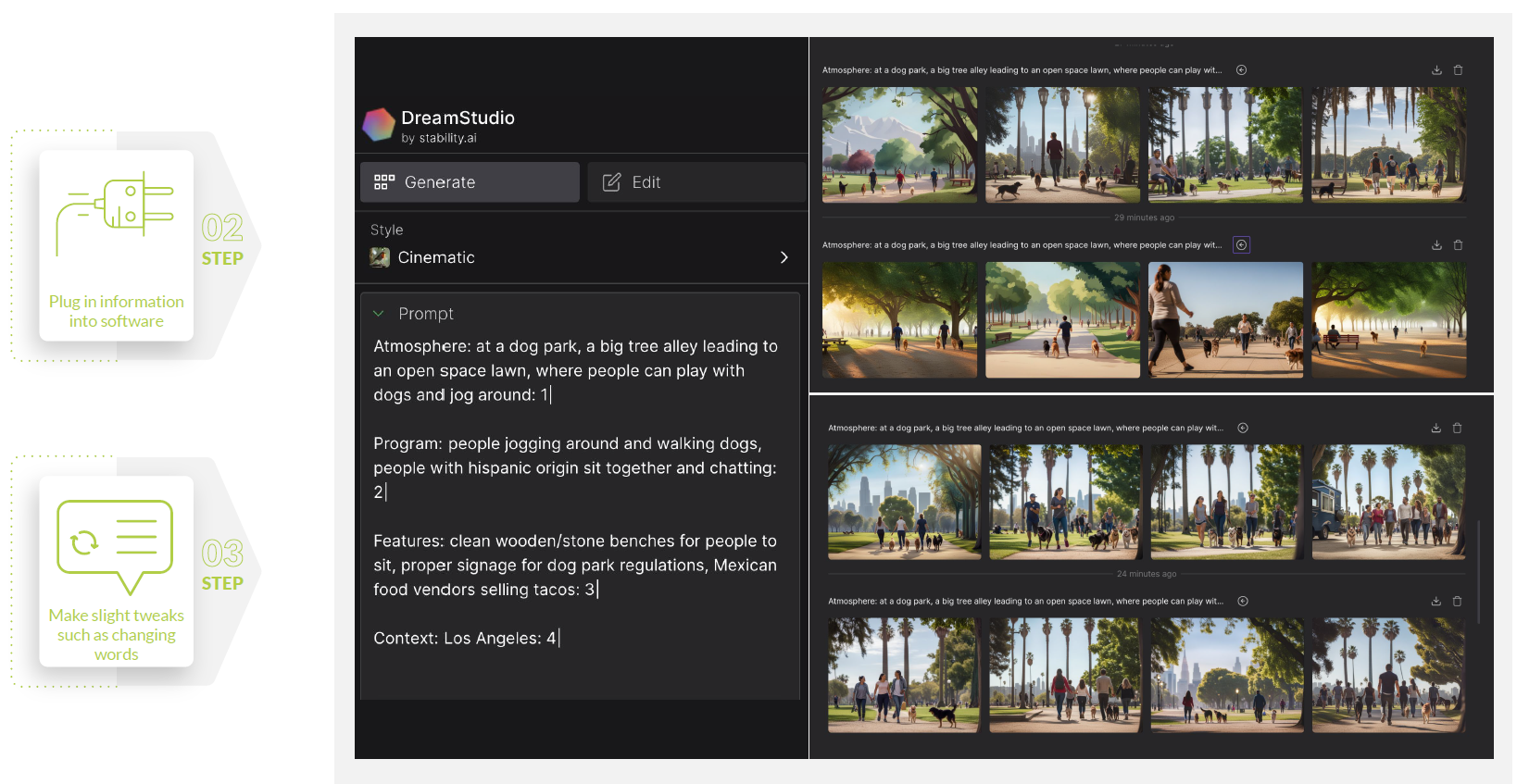}
         \caption{Stage 2 - Steps 2 and 3}
         \label{fig:Stage2(1)}
     \end{subfigure}
     \hfill
     \begin{subfigure}[h!]{0.95\textwidth}
         \centering
         \includegraphics[width=\textwidth]{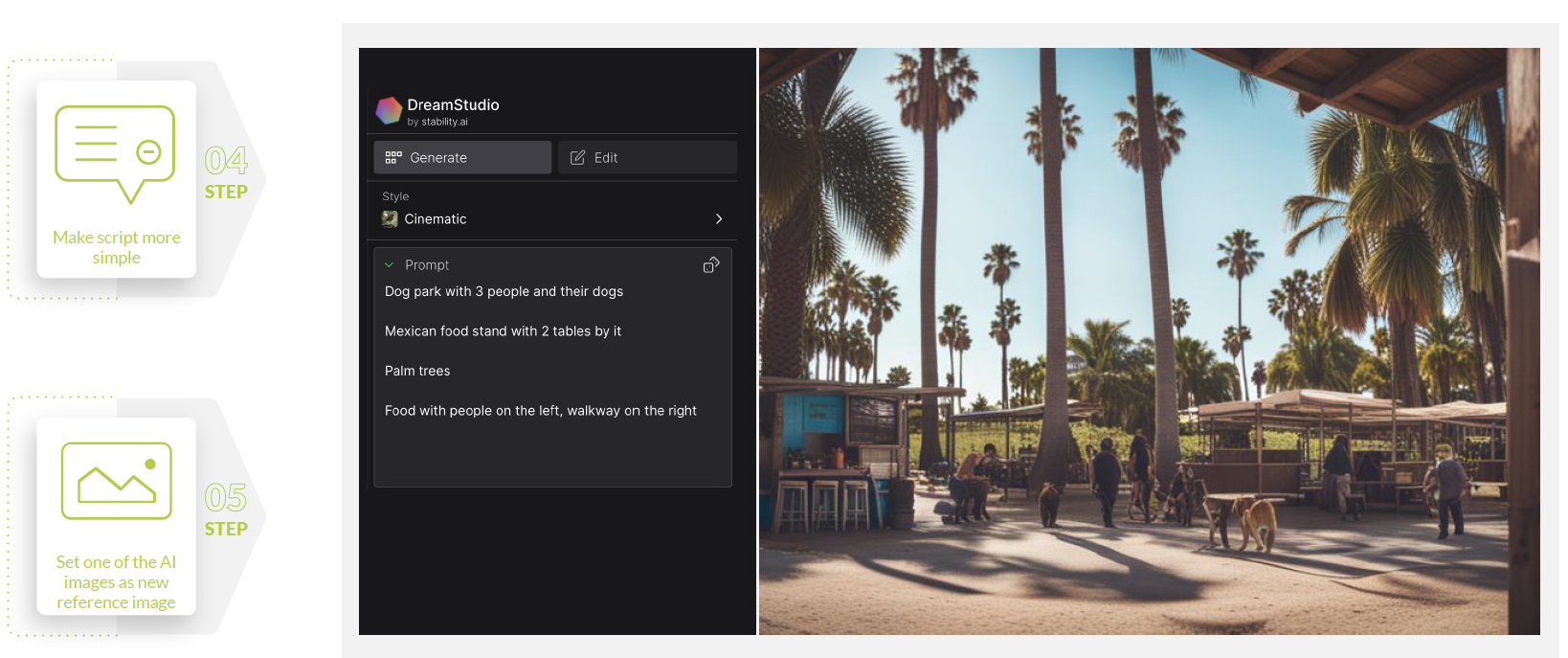}
         \caption{Stage 2 - Steps 4 and 5}
         \label{fig:Stage2(2)}
     \end{subfigure}
        \caption{Second Stage of Trials}
        \label{fig:SecondStage}
        \Description{Two figures show steps 2 to 5 to generate the image. Steps are (2) Plug the information into the software, (3) Make slight tweaks such as changing words, (4) Make script more simple, (5) Set one of the AI images as the new reference image}
\end{figure}
\end{document}